\documentclass[prc,twocolumn,showpacs,nofootinbib,aps,longbibliography,superscriptaddress]{revtex4-1}
\usepackage{natbib}
\usepackage[utf8]{inputenc}
\usepackage{color}
\usepackage{amsmath}
\usepackage{amssymb}
\usepackage{url}
\usepackage{graphicx}
\usepackage{amssymb}
\usepackage[dvipsnames]{xcolor}
\usepackage{appendix}
\usepackage[section]{placeins}
\usepackage{multirow}
\usepackage{hyperref}
\usepackage{mathrsfs} 
\usepackage{bm}
\usepackage{soul}
\usepackage{array}
\usepackage{lipsum}
\usepackage{relsize}
\usepackage{amsfonts}
\usepackage{comment}
\usepackage{outlines}
\usepackage[normalem]{ulem}
\usepackage{multirow}
\usepackage{orcidlink}

\usepackage{bbold}
\usepackage{booktabs}
\usepackage{tabularx}
\usepackage{xcolor}
\usepackage{enumitem}

\usepackage[ruled]{algorithm2e}
\DontPrintSemicolon
\SetNlSty{textbf}{}{:}
\SetAlgoNlRelativeSize{0}
\renewcommand{\vec}[1]{\mbox{\boldmath $#1$}}

\begin{document}

\title{Genetic Programming for the Nuclear Many-Body Problem: a Guide} 

\author{Illya Bakurov\,\orcidlink{0000-0002-6458-942X}}
\affiliation{Department of Computer Science and Engineering, Michigan State University, East Lansing, MI}

\author{Pablo Giuliani\,\orcidlink{0000-0002-8145-0745}}
\affiliation{Facility for Rare Isotope Beams, Michigan State University, East Lansing, Michigan 48824, USA}

\author{Kyle Godbey\,\orcidlink{0000-0003-0622-3646}}
\affiliation{Facility for Rare Isotope Beams, Michigan State University, East Lansing, Michigan 48824, USA}

\author{Nathaniel Haut\,\orcidlink{0000-0002-3989-6532}}
\affiliation{Department of Computational Mathematics, Science, and Engineering, Michigan State University, East Lansing, MI}

\author{Wolfgang Banzhaf\,\orcidlink{0000-0002-6382-3245}}
\affiliation{Department of Computer Science and Engineering, Michigan State University, East Lansing, MI}

\author{Witold Nazarewicz\,\orcidlink{0000-0002-8084-7425}}
\affiliation{Facility for Rare Isotope Beams, Michigan State University, East Lansing, Michigan 48824, USA}
\affiliation{Department of Physics and Astronomy, Michigan State University, East Lansing, Michigan 48824, USA}

\date{\today}

\begin{abstract}
Genetic Programming  is an evolutionary algorithm that  generates computer programs, or mathematical expressions, to solve complex problems. In this Guide, we demonstrate how to use Genetic Programming to 
develop surrogate models to mitigate the computational costs of modeling  atomic nuclei with ever increasing complexity. The computational burden escalates  when uncertainty quantification is pursued, or when observables must be globally computed for thousands of nuclei. 
By studying three  models in which the mean field depends on the total particle density self-consistently, we  show that 
by constructing reduced order models supported by Genetic Programming one  can speed up  many-body computations by several  orders of magnitude  with a negligible loss in accuracy. 
\end{abstract}

\maketitle

\section{Introduction}
\label{intro}
Computational models for many-body quantum systems, such as the atomic nucleus,  have quickly grown in complexity as theories are refined, more experimental data became available, and computers grew more powerful.  
In research areas  with practical applications, the demand for more accurate descriptions of physical phenomena, together with a push for predictions with well-quantified uncertainties, has driven various efforts to create surrogate models that can learn low-dimensional representations of the underlying system equations.

To overcome the dimensional barrier of nuclear models with many degrees of freedom,  we explore the application of Genetic Programming (GP) for obtaining reduced equations from data. 
GP is a type of Artificial Intelligence (AI) that falls under the broader category of evolutionary computation, which is a subset of AI inspired by biological evolution mechanisms such as natural selection and stochastic variation~\cite{b92_gp_koza,b98_gp_intro_banzhaf,b02_gp_foundations_langdon,b07_lineargp_brameier_banzhaf}.

Here we use GP in the context of the reduced basis method (RBM)~\cite{quarteroni2015reduced,hesthaven2016certified,bonilla2022training} that  works by identifying reduced coordinates as the amplitudes of a reduced basis informed by a small number of  high-fidelity (or full order model, FOM) evaluations of the system. 
Some of these developments follow a supervised machine learning (SML)~\cite{b16_deep_learning_goodfellow,b19_hands_on_ml_geron,b23_lectures_is_vanneschi_silva} approach whereby various algorithms are trained to reproduce the dynamics of the reduced coordinates as control parameters change. 

In this work, we explore GP to obtain the reduced equations from data. GP begins with a randomly generated initial \textit{population} of \textit{individuals} — each representing a model/equation of the control parameters in the reduced coordinates. The algorithm then simulates an evolutionary process over thousands of generations. During this process, individuals undergo selection based on their \textit{fitness}, which evaluates their effectiveness in capturing the underlying reduced dynamics of the system. Concurrently, stochastic modifications are made to their structure, similar to genetic mutations and crossover found in biological DNA. These modifications help to explore and exploit the search space by creating increasingly effective models that can discover simple yet accurate equations for the dynamics of the system in reduced coordinates as a function of the controlling parameters. With this approach we are able to create reduced-order models in a straightforward manner for describing many-body quantum systems that would have otherwise posed appreciable challenges with traditional approaches. 

This Guide is organized as follows.
The principles of GP are outlined in Sec.~\ref{method_gp}. Three illustrative many-body case studies are introduced in Sec.~\ref{sec: Quantum Systems}. Section
\ref{sec: RBM} describes the dimensionality reduction techniques employed.
In Sec.~\ref{sec: results} we present the results of our numerical tests  varying in functional complexity, regularization strength, and the inclusion of feature selection strategies. This comparative analysis helps to highlight the strengths and limitations of GP relative to conventional data-driven approaches. Finally, in  Sec.~\ref{sec: conclusions}, we close by presenting our conclusions and outlook for future directions.

\section{Genetic Programming}
\label{method_gp}

Genetic Programming (GP) is a type of Artificial Intelligence (AI) that falls under the broader category of Evolutionary Computation,
which is a subset of AI inspired by biological evolution mechanisms such as selection, mutation, and crossover to solve problems~\cite{b92_gp_koza,b98_gp_intro_banzhaf,b02_gp_foundations_langdon,b07_lineargp_brameier_banzhaf}.  GP is a bio-inspired search method making use of processes of stochastic variation and simulated natural selection.
GP is an extension of Genetic Algorithms to explore the search space of computer programs, that are usually dynamic in size, rather than searching for an optimal set of parameters as is typical in optimization problems. While Genetic Algorithms operate upon fixed-length strings of characters or numbers (genotype) which map to a solution (phenotype) of a given optimization problem, GP adjusts its complexity to the problem at hand, with a more powerful genotype-phenotype map that allows adaptation to a large class of problems.
Due to the high degree of flexibility of representations and modularity, GP has been extensively used in numerous machine learning (ML) domains: classification~\cite{c14_m3gp_ignalalli,c15_m3gp_munoz,a19_m4gp_lacava}, regression~\cite{c12_gsgp_moraglio,a19_gsgp_framework_castelli,c16_epsilon_lexicase_lacava}, feature engineering~\cite{a12_feature_construction_neshatian,a19_gp_hd_feature_construction_class_tran}, manifold learning~\cite{a22_manifold_lensen}, active learning~\cite{c23_active_learning_gp_haut}, ensemble learning~\cite{a21_stacking_gp_bakurov}, image enhancement, classification and segmentation~\cite{p94_recombination_selection_construction_tackett,c04_multi_class_image_zhang,c17_gp_skin_cancer_ain,a17_gp_image_class_iqbal, a23_ssnet_stack_gp_bakurov,c24_image_research_al_gp_haut,a22_elaine_correia,c21_gp_evolving_image_enhancement_correia}, and even generation of novel forms of AI~\cite{c15_sml_goncalves,b16_tpot_olson,c16_tpot_olson,a19_tpot_gp_olson,a20_automl0_real,arxiv23_discovering_adaptable_symbolic_kelly}. 

A common application of genetic programming is Symbolic Regression \cite{Keren2023,Angelis2023}, which is the learning of mathematical expressions to approximate some observed training data. There is recent evidence that Symbolic Regression can be a valuable tool for nuclear-physics modeling \cite{Munoz_2025,Cheng2024}.

There are many possible representations employed in GP but it is common to use expression trees in tree GP~\cite{b92_gp_koza}, linear sequences of instructions in linear GP~\cite{b07_lineargp_brameier_banzhaf}, and circuit-type graphs in Cartesian GP~\cite{b11_cgp_miller}. The choice of representation can have a significant impact on solution reachability since each representation requires different genetic operators (functions that modify the genotype) that impact how solutions transfer learned knowledge across generations \cite{c25_assumptions_model_Kotanchek}. 
In this work, we use tree GP and will refer to it as simply GP.

\begin{figure}[htb]
    \includegraphics[width=0.95\columnwidth]{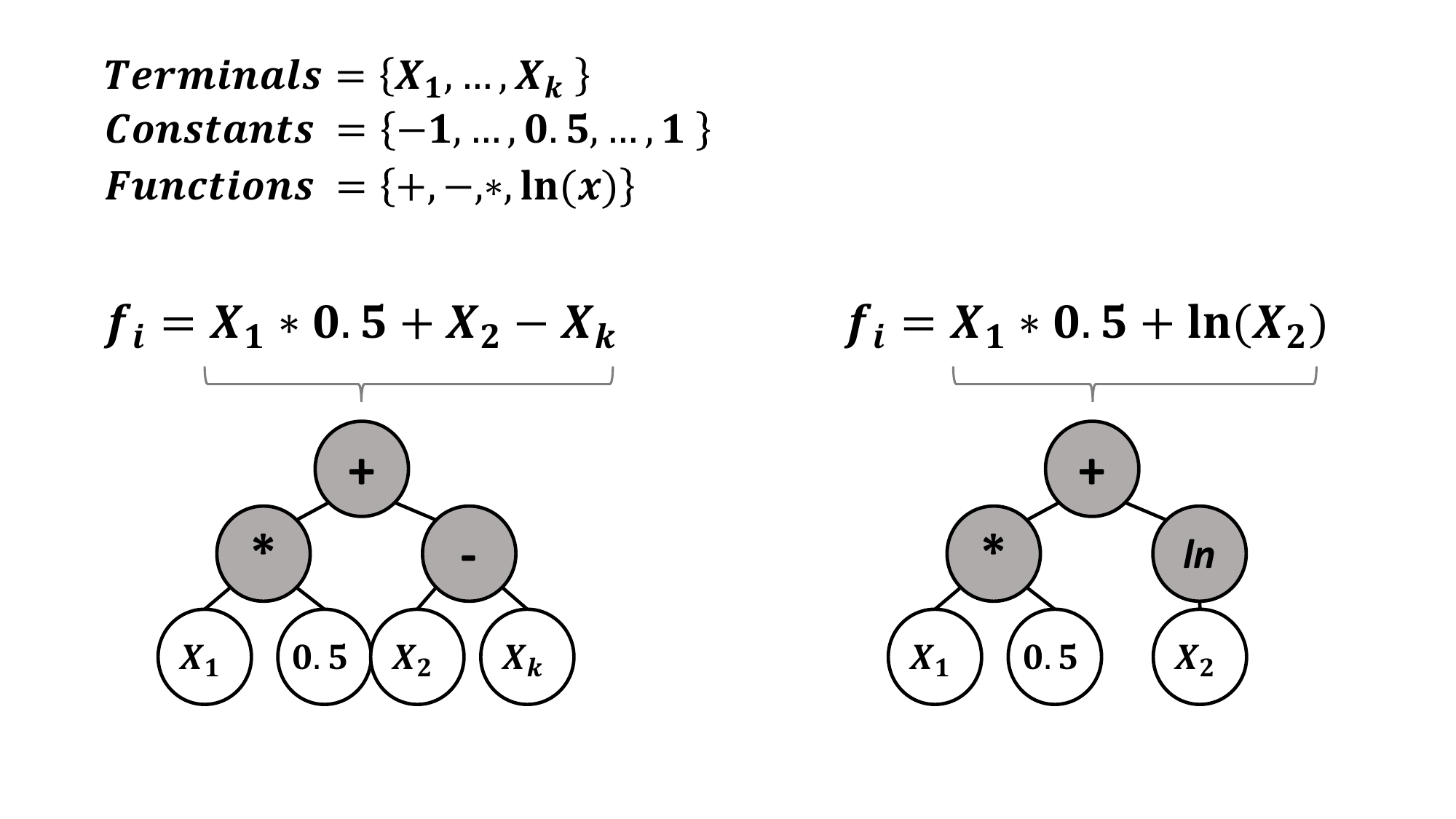}
    \caption{A diagram showing how mathematical functions can be represented as trees in Tree Genetic Programming. The gray nodes represent the mathematical functions of different arities, whereas the white nodes represent the input features and constants (aka terminal nodes).}
    \label{fig:tgp_example}
\end{figure}

In GP, the evolving programs are constructed by composing elements belonging to two specific, predefined, sets: a set of primitive functions $F$, which appear as the internal nodes of the trees, and a set of terminal input features $T$, and constant values $C$, which represent the leaves of the trees. In the context of Supervised Machine Learning (SML) problems, the trees encode symbolic expressions mapping inputs $T\ \cup \ C$ to outputs $Y$.
Figure~\ref{fig:tgp_example} provides a visualization of how GP-models can represent a mathematical function as a tree of functions, input features and constant values. Note that some functions are unary, while other are binary, which results in subtrees with one or two leaf nodes. In general, GP trees can use functions of any arity, although it is common to see just unary and binary operators. 

Typically, GP is used with the so-called subtree mutation and swap crossover~\cite{b92_gp_koza}.
The former randomly selects a subtree in the structure of the parent individual and replaces it with a new, randomly generated tree. The goal of mutation is to develop new skills in the population and refine existing ones (exploration). 
The swap crossover exchanges two randomly selected subtrees between two different parent individuals. The goal here is to transfer learned skills to new individuals to hopefully find improved combinations of existing skills in the population (exploitation). The quality of an individual is assessed using a \emph{fitness} function. The fitness function is problem specific, but should capture how well an individual accomplished the task. For example, in symbolic regression, root-mean-squared error and $R^2$ are often used as  fitness functions. 
Algorithm~\ref{alg:gp} provides pseudo-code for a typical GP algorithm. In the algorithm, elitism refers to the process of preserving the best individual(s) with respect to the fitness function across generations. This ensures that the algorithm does not lose progress, and retains a good-performing individual.

\begin{algorithm}[tbh]\label{alg:gp}
\caption{Pseudo-code for the typical GP algorithm.}

\textbf{1:} Create an $n$-sized initial population of random individuals $P$\;

\textbf{2:} Repeat until matching a termination condition (e.g.\ number of generations):\;

\hspace*{0.2em}\textbf{(a):} Calculate fitness $\forall$ individuals $i$ in $P$\;

\hspace*{0.2em}\textbf{(b):} Initialize child population $P'$\;

\hspace*{0.2em}\textbf{(c):} Repeat until $P'$ contains $n$ individuals:\;

\hspace*{1em}\textbf{(i):} With probability $P_c$, choose crossover (transfer)\\
\hspace*{3.3em}operator; else choose mutation ($1 - P_c$)\;

\hspace*{1em}\textbf{(ii):} Select 2 individuals using a selection algorithm\;

\hspace*{1em}\textbf{(iii):} Apply the operator to the selected individuals\;

\hspace*{1em}\textbf{(iv):} Insert the resulting individuals into $P'$\;

\hspace*{0.2em}\textbf{(d):} Replace $P$ with $P'$ (copy elites from $P$ if elitism\\
\hspace*{2.6em} is used)\;

\textbf{3:} Return the best individual(s) in $P$\;

\end{algorithm}

In a ML context, GP provides users a  different experience in terms of transparency and interpretability of the final models when compared to more ``black-box" models of, e.g., deep artificial neural networks. This is because GP typically outputs symbolic expressions or structures representing programs, which are generally interpretable and can be analyzed by the researcher. By manipulating discrete structures, the inner workings of a GP-based model are clearer, and the relationships between features and predictions are explicitly defined. As a result, evolved solutions can offer valuable insights into how the model arrives at its predictions, allowing users to understand the algorithmic decisions and relate this to their domain-knowledge. 

Beyond the models just being symbolic, GP has the benefit that it evolves a \emph{population} of models rather than just a single solution. Having access to a final population of models unlocks a lot of opportunities that are unavailable to other machine learning workflows. For example, while it is customary to just pick the most accurate model from the population on the training set, it is possible to explore many of the candidate models and use human expertise to select from those models the one, or multiple, that best align with domain knowledge. Rather than picking one model from the evolved population, there are methods used in GP to select a model ensemble to be used in multi-modeling approaches \cite{kotanchek2008trustable}. GP ensembles can leverage the diversity of patterns to improve stability in deployment and also mitigate single-model risks. 

The populations of solutions can also be studied to gain insight into the system of study. For example, it is possible to detect patterns regarding which features are common in the population, or which sets of features commonly appear together in models. This can provide researchers with key insights into which features provide predictive synergies. 

Although GP is not the only ``white-box" approach in the spectrum of ML methods, it allows for unprecedented flexibility of representations and modularity, making it suitable for numerous tasks.
The area of explainable AI receives consistently more attention from both practitioners and researchers~\cite{a21_role_exai_markus,a23_exai_ethics_pekka} and GP has gained popularity where human-interpretable solutions are paramount. Real-world examples include medical image segmentation~\cite{a23_kartezio_med_image_segment_cortacero}, prediction of human oral bioavailability of drugs~\cite{a07_gp_pharma_archetti,a12_bloat_free_gp_bioav_silva}, skin cancer classification from lesion images~\cite{a22_gp_skin_cancer_qurrat}, and even conception of models of human visual perception~\cite{a23_friqa_gp_bakurov}. 
Besides being able to provide interpretable models, there is evidence that GP can also help to unlock the behavior of black-box models~\cite{c19_gp_insite_blackbox_evans,a23_ssnet_stack_gp_bakurov}. 

A distinctive advantage of GP, is the flexibility to use any optimization objective or to perform multi-objective optimization~\cite{a19_m3gp_multi_trans_munoz,a22_semantics_moo_gp_galvan,c24_m6gp_batista,Kotanchek2007}. For example, many deep learning approaches require gradient information making them unsuitable for problems where derivatives are difficult to compute or do not exist~\cite{b16_deep_learning_goodfellow}.
In contrast, GP can use any objective function as long as there is some way to distinguish between more and less \textit{fit} candidate-solutions (individuals) in a population.

ML methods for regression, such as LR, artificial neural networks~\cite{b16_deep_learning_goodfellow}, and support vector machines~\cite{b08_svm_vapnik}, typically rely on error-based metrics (penalty function)  such as the mean squared error to guide the model optimization. However, in the context of GP, such penalty functions have been found to be less effective at guiding the search toward solutions that both (i) exhibit the desired geometric structure of the response surface in the input space, and (ii) closely fit the target data~\cite{c03_gp_interval_scaling_keijzer,c07_mo_gp_generalization_vanneschi}.
To mitigate this issue, Ref.~\cite{c03_gp_interval_scaling_keijzer} proposed to estimate the optimal linear transformation (slope and intercept) of the GP-generated expression with respect to the target outputs, allowing the evolutionary process to focus primarily on discovering the correct functional form or \textit{shape} of the model. Although effective, this approach requires recomputing the scaling parameters at every generation for each individual. 
Later it was understood that using the coefficient of determination ($R^2$, i.e., squared Pearson's correlation) removes the need for explicit linear scaling and translation, as it directly identifies models that match the shape or form of the target function~\cite{c05_pareto_tournament_gp_smits,c07_trustable_pareto_tournament_gp_kotanchek,c09_trustable_gp_targeted_collection_kotanchek}.

Recently, Ref.~\cite{c23_gptp_r2_haut} emphasized that using $R^2$ as the fitness function in GP for SR can lead to significantly better generalization than standard RMSE (without scaling), particularly when training on small datasets. Importantly, this improvement is achieved without the need for computationally expensive linear scaling at each generation. Instead, scaling is applied only once as a post-processing step after the evolution. However, their experiments also revealed that while $R^2$ performs well under low-noise conditions, its advantage diminishes as noise levels increase, eventually degrading to the performance level of RMSE or worse. This suggests that $R^2$ is less robust to noise, limiting its effectiveness in high-noise regimes.
Later, more evidence on the utility of $R^2$ to relieve GP from learning coefficients in SR was provided in~Refs. \cite{c23_relieving_chen,c25_lexicase_bakurov}. In particular, in~\cite{c25_lexicase_bakurov} it has been found that $R^2$ improves generalization of evolved GP models regardless of the selection scheme and pressure. 
%
%
\subsection{GP Software}
\label{method_gp_illustration}
Several open-source GP libraries are available, each addressing different audiences, needs, and perspectives with their GP implementations. 
GPLearn offers a scikit-learn-compatible API in Python for applying tree-based GP in SML problem-solving~\cite{gplearn}.
Distributed Evolutionary Algorithms in Python (DEAP) is a Python-based framework for evolutionary algorithms, including tree-based GP~\cite{DEAP_JMLR2012}. Moreover, it offers parallel processing capabilities and its modular implementation is suitable for rapid prototyping.
PushGP evolves programs in the stack-based Push language, designed to be used as the programming language within which evolving programs are expressed~\cite{a02_pushgp_spector}. 
General Purpose Optimization Library (GPOL) provides a unified Python interface for a wide range of algorithms (including tree- and linear-based GP), supports CPU/GPU computations and batch processing~\cite{a21_gpol_bakurov}. 
KarooGP~\cite{karoogp} and TensorGP~\cite{baeta2021tensorgp} utilize TensorFlow to accelerate GP through vectorized operations on GPU, enhancing performance on large-scale problems.
Waikato Environment for Knowledge Analysis (WEKA) is a Java-based software that provides a collection of several machine learning algorithms, including GP, for data mining and knowledge discovery~\cite{a09_weka}. 
HeuristicLab is a C\# framework offering a wide range of algorithms for solving optimization problems, including GP, with a user-friendly GUI for experiment design and analysis~\cite{hlab_Wagner2014}.
In this paper, we rely on a commercially available system, called DataModeler~\cite{datamodeler}. It offers a user-friendly GUI and, most importantly, implements several SOTA GP techniques. The latter feature is one its biggest attractions, as open-source implementations frequently use standard methods, which might result in under-performance. Moreover, the software provides discounted licenses for universities, facilitating its adoption in academic research.

%
%
\subsection{GP Illustration}
\label{method_gp_illustration}

To illustrate the application of GP in a Symbolic Regression task, let's take 
Equation~II.3.24 from the Feynman Symbolic Regression Database \cite{FeynmanSRData} that relates  flux heat ($Flux$) to the power source ($Pwr$):
\begin{equation}
\label{eq:feynman53}
    Flux = Pwr/(4*\pi*r^2).
\end{equation}

Taking into account this equation, we built a data set with 6 variables and 100 data records. Each data record was pulled from a uniform distribution. Two of the variables are the real inputs, $Pwr$ and $r$, two of the variables, $rand1$ and $rand2$, are randomly generated, and two of the features are noise-corrupted versions of $Pwr$ and $r$, which we will call $PwrNoise$ and $rNoise$. The first four features ($Pwr$,$r$,$rand1$, and $rand2$) are limited to values between 1 and 3. Noise-corrupted versions are generated with 50\% noise where $NoiseData=Data + \epsilon$ with $\epsilon$ being uniformly distributed $\pm25\%$ the range of $Data$. Random and noisy features were included here to illustrate the ability of GP to perform automatic feature selection and focus on only the most informative (noiseless) features. Once the data set is setup, we can proceed with applying GP to discover the functional relationship. 

The first step in using GP is to choose the operator set (choice of mathematical operators). When setting up the operator set, the goal is to include enough operators 
so that it is useful, but not too many such that the search space becomes excessively large. It is always possible to refine the strategy based on the performance of the current set, thus, it is not detrimental to include too many or too few in the first attempt. In this case, we include the following operators, which happen to be the default set in DataModeler~\cite{datamodeler}: $+$, $-$, $*$, $/$, $negate$, $x^2$, $sqrt$, $inverse$, $x^n$, and $continuedFraction$. 

\begin{figure}[ht!]
    \centering
    \includegraphics[width=\linewidth]{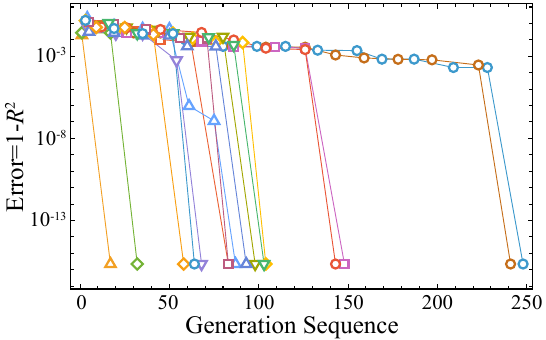}
    \caption{GP search progress over generations. The various colors represent the independent parallel searches. In this case, we ran on 16 cores so there are 16 parallel searches. The symbols represent the best fitness of all models in a specific generation. 
    }
    \label{fig:gpSearch}
\end{figure}

Once the operator set is chosen, we can set the termination condition, typically either run-time or number of generations. In this case we chose run-time, as it is typically more useful in application, with generation count more interesting for GP theory.
In this example, we set run-time to 10 seconds, after which the search terminates. This time window controls the complexity of the equations that the GP can find, yet in practice this time-controlled complexity cap will vary with the particular machine used.  
We can also set the number of parallel searches that we want to use. Typically, we set this to the number of cores available on the machine. In this case, that means 16 cores, i.e.,  16 parallel searches. The search trace
on this example problem is shown in Fig.~\ref{fig:gpSearch}. We can see that all of the parallel searches found a solution within the 
time frame limit.

\begin{figure}[ht!]
    \centering
    \includegraphics[width=\linewidth]{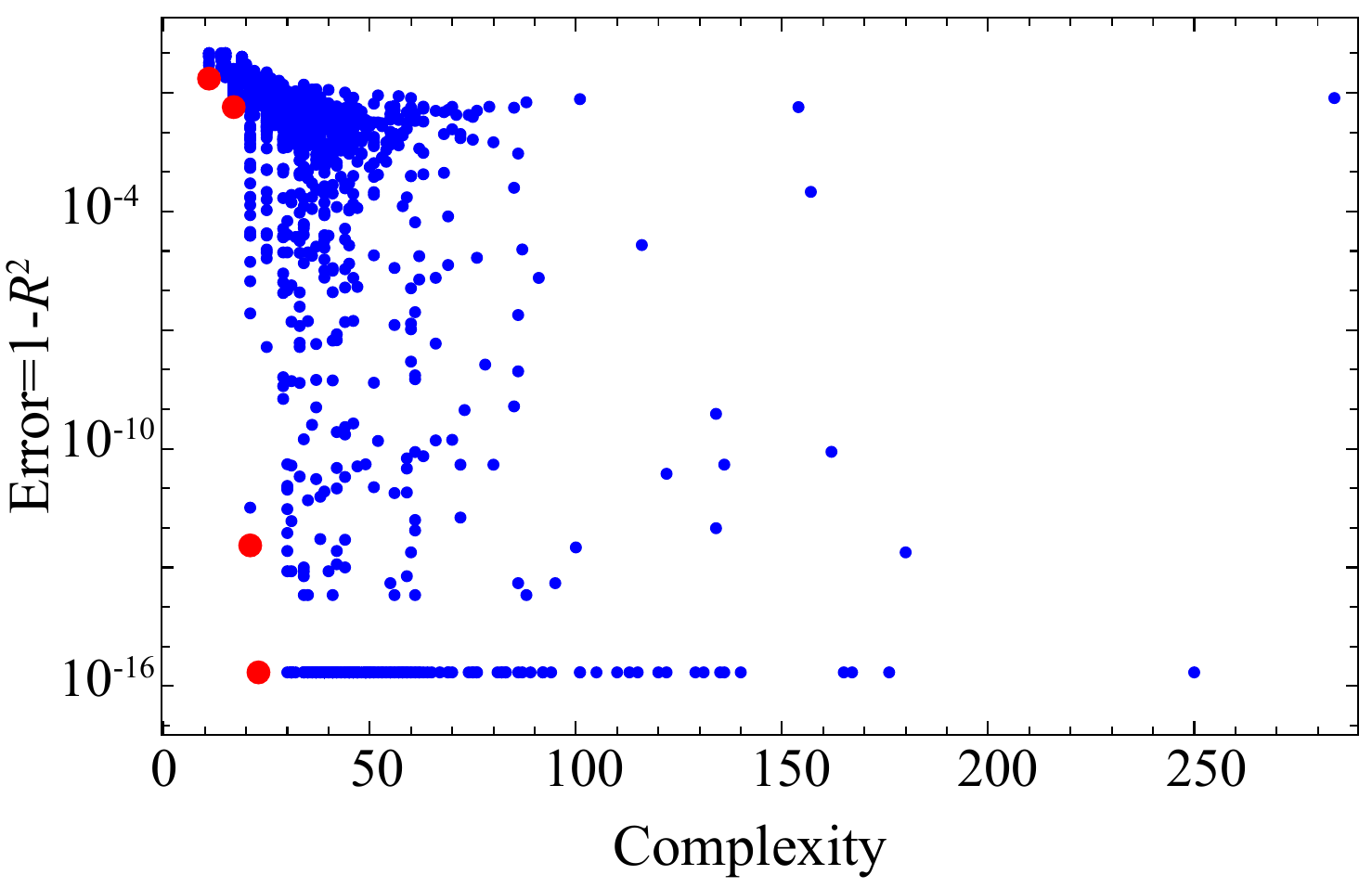}
    \caption{The Pareto front of GP models with respect to complexity and accuracy for illustration. The best model found is displayed and shows that the correct model was found by search. The red points represent the Pareto front of models. } 
    \label{fig:gpFront}
\end{figure}

\begin{figure}[ht!]
    \centering
    \includegraphics[width=\linewidth]{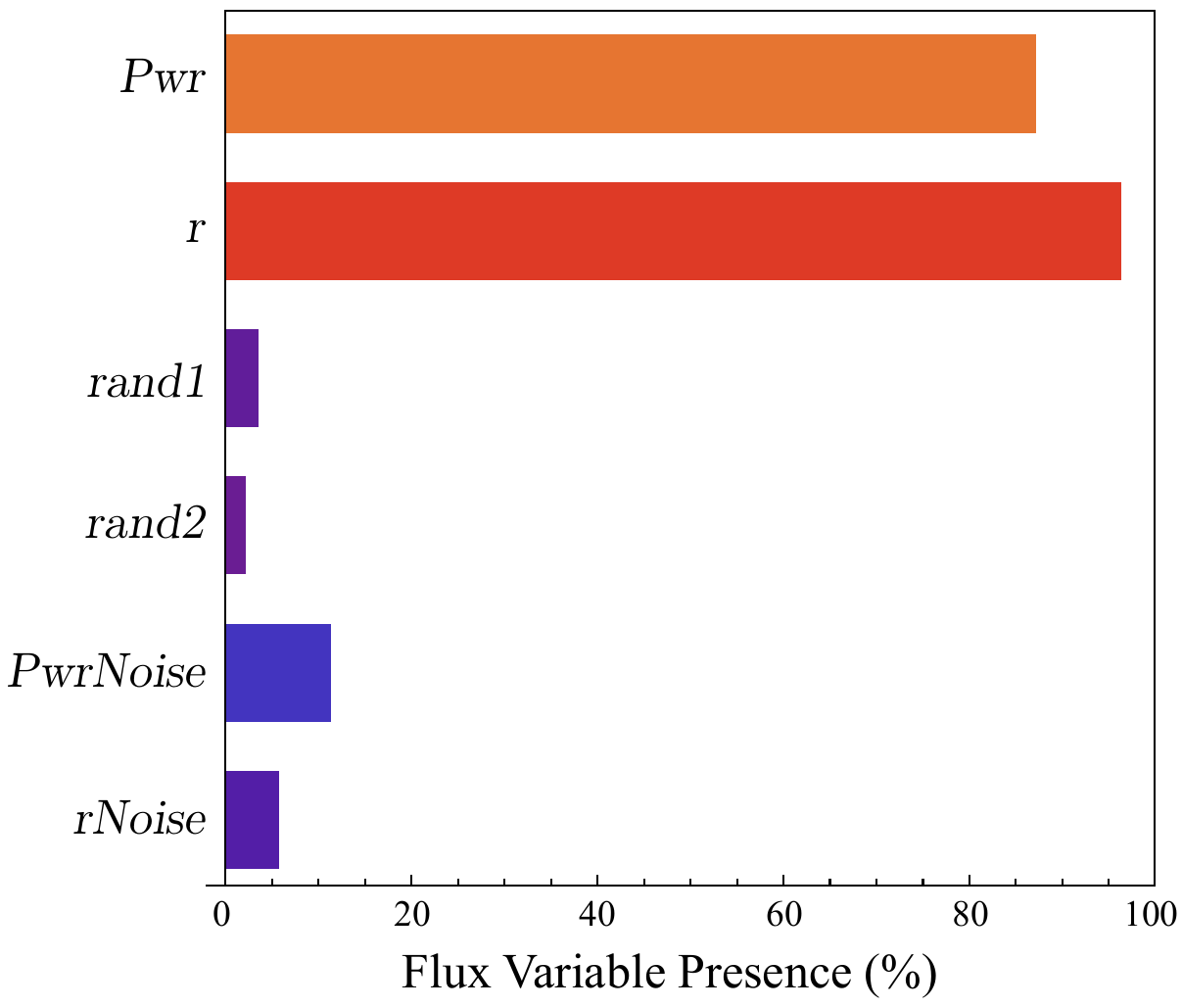}
    \caption{The percentage of models that include each variable from the sample data set. Showcasing GP's ability to focus on most informative features.} 
    \label{fig:gpVars}
\end{figure}

Once the GP search is complete, we can extract the evolved models to select a model, or a model ensemble, as well as explore properties of the evolved populations. Figure~\ref{fig:gpFront} shows the Pareto front plot of the evolved population along with the best model found:
\begin{equation}
    Flux = 0.07957747154594767 \frac{Pwr}{r^2}+9.34\cdot 
    10^{-18},
\end{equation}
which is the correct model within machine-precision errors.
The red points represent the Pareto front \cite{Kotanchek2007}(non-dominated models with respect to complexity and accuracy), while the blue points represent the rest of the models returned by the search. As discussed, the population can also be studied directly to reveal important information. For example, we can look at the percentage of models that include each of the variables in the data set, as shown in Fig. \ref{fig:gpVars}. This shows that the GP search was able to correctly identify the useful variables and mostly ignored the noise-corrupted and non-informative variables. It is important to note that the best model found contains only noise-free features. Since models in the population are randomly mutated, it is expected that some models may contain uninformative features in each generation, but the fraction of models with those uninformative features is small compared to the fraction of models using the informative features. If exploring a problem with an unknown form, this information could reveal which variables are key to understanding the system of study. 

Although there is much more flexibility and features of GP, this illustration provides a simple and direct overview of how GP can be applied to a sample data set while providing some context for how it is applied  in the rest of this work. 

%
%
\subsection{Linear Regression}
\label{method_lr}

We benchmark the performance of GP as an alternative to build the reduced expressions for surrogate models with that of Linear Regression (LR), one of the most widely used supervised machine learning methods for regression, known for its simplicity, interpretability, and efficiency. LR assumes a linear relationship between input variables/features ($x$) and the output target variable ($y$), expressed as a weighted sum of terms: $y = \beta_0 + \beta_1 x_1 + \beta_2 x_2 + \ldots + \beta_n x_n$. Here, the coefficients $\beta_k$ are learnable parameters associated with the intercept and $k^{th}$ input feature. \textit{Learning} of parameters occurs by minimizing the sum of squared residuals (i.e., difference between the observed target values in the training dataset, and the values predicted by the linear model). 

However, LR exhibits a fundamental limitation: it relies upon a fixed additive structure where the functional form of the input features and their interactions must be pre-defined by the user. This constrains LR's ability to accurately model complex nonlinear relationships or discover concise and appropriate functional representations of the data. GP, in contrast, explores and even composes diverse mathematical transformations and interactions without rigid assumptions. 

To overcome the limitation of LR and make it more expressive, we designed a custom feature transformation pipeline that augments the original input space with various functional forms (e.g., powers, roots, logarithms, exponentials) and interaction terms (products and quotients). Recognizing the risk of overfitting due to the inflated feature space and fixed data size, we incorporate regularization techniques (such as L1~\cite{a96_lasso_tibshirani}, L2~\cite{Fridmann:2005}, ElasticNet~\cite{a05_elasticnet_zou}, and up to two consecutive feature selection steps. The method is based on the Extremely Randomized Trees ensemble~\cite{a06_extra_trees_geurts} - a method that combines decision-tree regression models similar to Random Forests~\cite{a01_rf_breiman}, but with stronger randomization of the split point when inferring trees. Specifically, instead of exhaustively enumerating all possible splits to choose the one that minimizes a given loss function, it randomly chooses a set of splitting points and then selects the best from those. This strategy was shown to report a better generalization ability and faster run-times when compared to Random Forests. Moreover, it was found to be particularly useful when dealing with high-dimensional and noisy data~\cite{a06_extra_trees_geurts} -- the  reason why we employed Extremely Randomized Trees in the first step of our feature selection pipeline. 
The second feature selection step uses LR itself for refinement as it keeps the input features with above-median absolute coefficients. This allows us to focus on \textit{strong} predictors based on linear influence.

Both feature transformations, regularization and feature selection steps are treated as hyperparameters of the pipeline, which are tuned by means of cross-validation. This strategy not only enhances LR's ability to model non-linearities but also ensures a fairer comparison with GP. Full implementation details and reference to the source code are provided in Appendix~\ref{sec:appendix}.

\section{Many-Body Case Studies}\label{sec: Quantum Systems}

\subsection{Modified Gross-Pitaevskii equation}

The first case study considered is a modified version of the one-dimensional modified Gross-Pitaevskii equation (MGPE),
 which is a nonlinear Schr\"odinger equation originally introduced to describe the ground state of identical bosons (Bose-Einstein condensate). The MGPE used in this study can be written as:
\begin{equation}\label{eq: BGG}
   \left[-\frac{d^2}{dx^2} + \kappa x^2 + q\rho(x)^\sigma\right] \phi_i (x) = E_i \phi_i (x).
\end{equation}
The mean-field Hamiltonian for a single-particle wave function $\phi_i$ consists of a trapping term proportional to $x^2$, as well as an interaction term depending on the total particle density $\rho$.
In its original form~\cite{dalfovo1999theory}, the particle density appears linearly, i.e., $\sigma=1$.  In this work, we modified the original equation to include more general, even fractional, powers of the density, resulting in three total control parameters $\alpha=\{\kappa,q,\sigma\}$. 
We also extended the MGPE  to fermions. In this case, the particle density can be written as:
\begin{equation}\label{eq: density}
    \rho(x) = \sum_i^N |\phi_i(x)|^2,
\end{equation}
where  $N$ is the number of fermions.

These extensions stem from our desire to construct a simple model with a parametric dependence that mimics the fractional power density terms present in some nuclear models. As such, Eq.~\eqref{eq: BGG} should be viewed as an illustrative case for the non-affine and nonlinear parametric dependence that challenges the application of some reduced-order model techniques~\cite{quarteroni2015reduced,bonilla2022training}.

\subsection{Skyrme  Density Functional Theory}

The second case study is the description of the complex nucleus $^{48}$Ca using nuclear  Density Functional Theory (DFT).
DFT is widely used in nuclear physics to describe complex atomic nuclei in a self-consistent framework. In nuclear DFT, the nucleons (protons and neutrons) interact through an effective interaction in the medium described by the energy density functional (EDF) of  densities and currents.

The first DFT framework used in this study is a non-relativistic Skyrme-DFT (SDFT). 
In the simplest time-even  variant, one considers local particle ($\rho$), kinetic ($\tau$), spin-current
tensor ($\mathbb{J}$), and spin-orbit current ($\vec{J}$) densities (see Refs.~\cite{Bender2003,nucleardft} for more details). Each of these densities is a function of the spatial coordinate $\vec{r}$, e.g., $\rho=\rho(\vec{r})$.
In the following, this set of densities is denoted as $\{\rho\}\equiv\{\rho, \tau, \mathbb{J},\vec{J}\}$. 
The total energy of the nucleus  can be written as a functional of 
$\{\rho\}$: 
\begin{equation}
E=E[\{\rho\}]=\int {\cal H}(\vec{r})d^3{\vec r},
\end{equation}
where $\cal H(\vec{r})$ is the Hamiltonian density.

The Skyrme EDF \cite{vautherin1972,engel1975,Bender2003} can be expressed through 
densities $\{\rho\}$ as
\begin{equation}
{\cal H}(\vec{r}) = {\cal H}_0(\vec{r})+ {\cal H}_1(\vec{r}),
\end{equation}
with
\begin{eqnarray}
\label{eq:edf}
    {\cal H}_t(\vec{r})&=& C_t^{\rho}\rho_t^2
    + C_t^{\rho\Delta\rho}\rho_t\Delta\rho_t
    + C_t^{\tau}\rho_t\tau_t
    + C_t^{   J} \mathbb{J}_t^2 \nonumber \\
    &+& C_t^{\rho\nabla J}\rho_t\vec{\nabla}\cdot\vec{J}_t,
\end{eqnarray}
where the isospin index $t$ labels isoscalar ($t = 0$) and isovector
($t = 1$) densities.  The Skyrme EDF coupling constants (here: controlling model parameters) are denoted as $\{ C_t^{\rho},C_t^{\rho\Delta\rho} ,C_t^{\tau} , C_t^{   J},  C_t^{\rho\nabla J}\}$. This functional is supplemented by the Coulomb term that describes the electromagnetic interaction, for which the coupling constant is fixed.

By varying the energy functional $E[\{\rho\}]$ with respect to the densities 
$\{\rho\}$  one obtains the mean-field, or Hartree-Fock,  Hamiltonian $\hat{h}$.
In the absence of nucleonic pairing, the single-nucleonic wave functions  are obtained by solving  the self-consistent Hartree-Fock eigenproblem:
\begin{equation}\label{eq:HF}
    \hat{h}[\{\rho\}]\phi_i = e_i\phi_i,
\end{equation}
which is a coupled non-linear  
integro-differential equation as the densities  $\{\rho\}$ are built from single-particle wave functions $\phi_i$ as in Eq.~(\ref{eq: density}).

\subsection{Covariant Nuclear Density Functional Theory}

The second DFT framework we consider is the covariant DFT (CDFT) approach, also known as relativistic mean-field theory. The basic assumptions of CDFT are similar to those of the Skyrme-DFT. Since the underlying formalism is relativistic, 
the Hamiltonian density is replaced by
the Lagrangian density whose effective degrees of freedom are
nucleons and mesons ~\cite{Serot1992,Ring1996,Bender2003,Schunck2019}.
This Lagrangian density  consists of a
nucleon-nucleon interaction mediated by  various mesons alongside non-linear meson 
interactions\,\cite{Boguta:1977xi,Mueller:1996pm}:
\begin{eqnarray}
{\mathscr L}_{\rm int} &=&
\bar\psi \left[g_{\rm s}\phi   \!-\! 
         \left(g_{\rm v}V_\mu  \!+\!
    \frac{g_{\rho}}{2}{\mbox{\boldmath$\tau$}}\cdot{\bf b}_{\mu} 
                               \!+\!    
    \frac{e}{2}(1\!+\!\tau_{3})A_{\mu}\right)\gamma^{\mu}
         \right]\psi \nonumber \\
                   &-& 
    \frac{\kappa}{3!} (g_{\rm s}\phi)^3 \!-\!
    \frac{\lambda}{4!}(g_{\rm s}\phi)^4 \!+\!
    \frac{\zeta}{4!}   g_{\rm v}^4(V_{\mu}V^\mu)^2 +  \nonumber \\
    &-& 
   \Lambda_{\rm v}\Big(g_{\rho}^{2}\,{\bf b}_{\mu}\cdot{\bf b}^{\mu}\Big)
                           \Big(g_{\rm v}^{2}V_{\nu}V^{\nu}\Big),                                           
 \label{LDensity}
\end{eqnarray}
where the nucleon fields are denoted by $\psi$,  meson fields  by $\phi$, $V$, and $\boldsymbol{b}$, and the Lagrangian coupling constants (controlling  parameters) are $\{g_{\rm s}, g_{\rm v},g_{\rho},
\kappa,\lambda,\Lambda_{\rm v},\zeta\}$. From this Lagrangian density, coupled equations can be derived for the meson fields (the Klein-Gordon equations) and the nucleonic wave functions $\phi_i$ (the Dirac equation).  For links between CDFT and SDFT, see  Ref.\,\cite{Bender2003}.

\subsection{Parameter Ranges}

\begin{figure}[htb]
  \includegraphics[width=1\columnwidth]{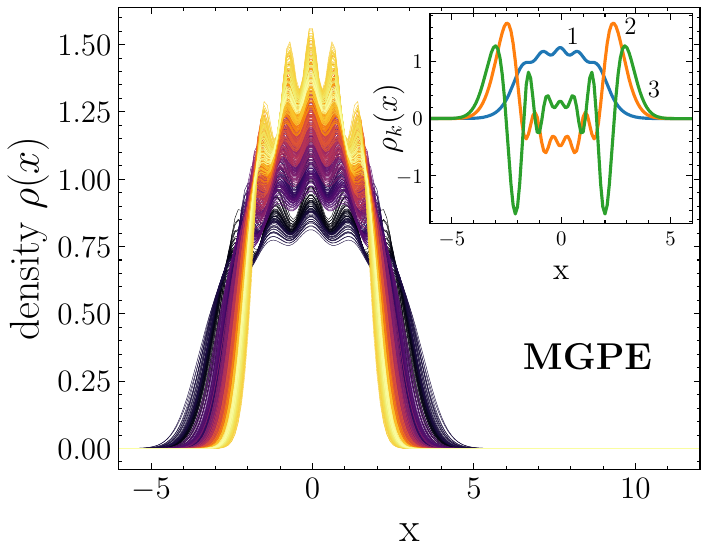}
    \caption{Densities $\rho(x)$ obtained in MGPE  for the 492 training evaluations across the parameters listed in Table~\ref{tab:params}. The inset shows the three principal components defined in Eq.~\eqref{eq: rho expansion} for $k=1,2,3$. All quantities are in arbitrary units. }
    \label{fig:BGGDensities}
\end{figure}

Table~\ref{tab:params} shows the ranges of parameters explored in the three case studies. 
As an illustrative example, Fig.~\ref{fig:BGGDensities}, pertaining to the MGPE,  shows the density $\rho(x)$ for a system with $N=5$ for various values of the control parameters. 

\begin{figure}[htb]
    \includegraphics[width=1\columnwidth]{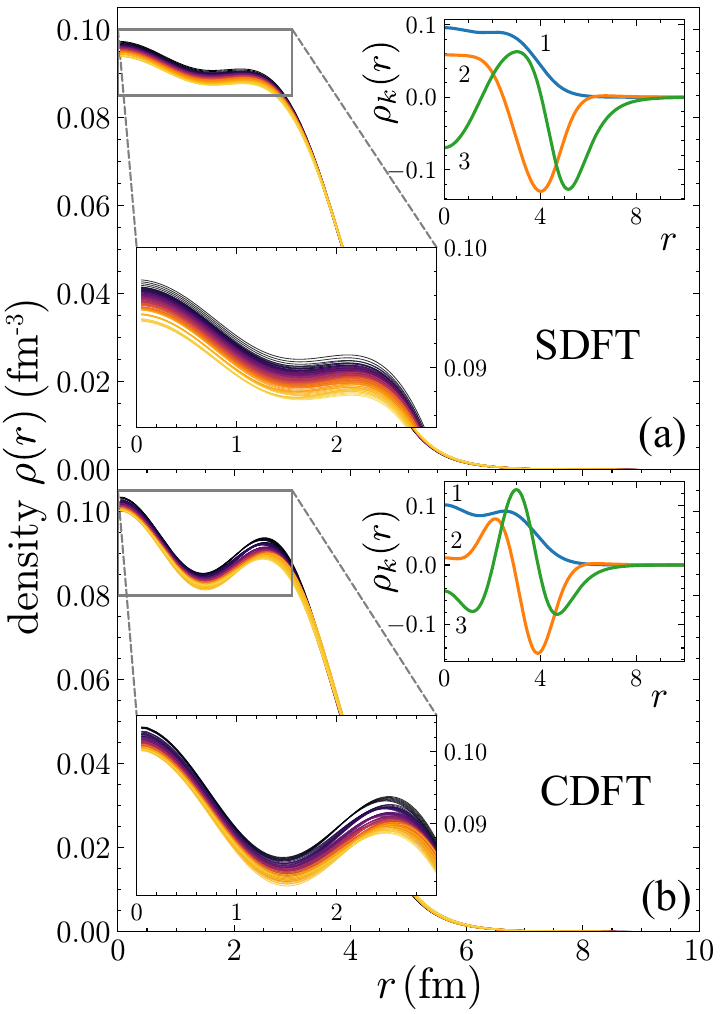}
    \caption{Neutron densities $\rho(r)$ for $^{48}$Ca calculated in  (a) SDFT   and (b) CDFT within the parameter ranges listed in Table~\ref{tab:params}. For both cases, 49 training and 49 testing parameter sets were evaluated, but 9 of the CDFT testing evaluations were excluded due to convergence issues. The inset shows the three principal components (in fm$^{-3}$) defined in Eq.~\eqref{eq: rho expansion} for $k=1,2,3$. }
    \label{fig:DFT Densities}
\end{figure}

\begin{table}[htb]
   \caption{Ranges of the parameters used to train and test the reduced order model in the three case studies considered.
   The MGPE parameters are dimensionless. The DFT parameters $J$ and $L$ are in MeV.
   492 evaluations of the MGPE  were used for training the reduced models, and 212 evaluations were sampled across the same parameter range for testing. 49 evaluations of the SDFT and CDFT models were used for training the reduced models, and 49 for testing them in both cases, sampled from the ranges specified in the table under the subscripts ``train" and ``test", respectively.}
  
    \label{tab:params}
 \begin{ruledtabular}   
    \begin{tabular}{c|c|c}
       {\bf Case Study} & {\bf Parameter} & {\bf Range} \\
        \hline
        \multirow{3}{*}{MGPE}
            & $\kappa$ & [0.5, 3.0]$_\text{train \& test}$\\
            & $q$ & [$-$2.0, 2.0]$_\text{train \& test}$ \\
            & $\sigma$ & [0.5, 3.0]$_\text{train \& test}$ \\
        \hline
        \multirow{2}{*}{SDFT}
            & $J$  & [25, 35]$_\text{train}$ , [20, 43]$_\text{test}$ \\
            &  $L$ & [20, 60]$_\text{train}$ , [10, 90]$_\text{test}$ \\
        \hline
        \multirow{2}{*}{CDFT}
            & $J$  & [30, 40]$_\text{train}$   [28, 50]$_\text{test}$\\
            & $L$  & [50, 130]$_\text{train}$ ,[40, 160]$_\text{test}$ \\
        \bottomrule
    \end{tabular}
\end{ruledtabular}
\end{table}

The control parameters in both DFT frameworks, for example $C_t^{\rho}$ in Eq.~\eqref{eq:edf} and $g_{\rm s}$ in Eq.~\eqref{LDensity}, determine how the solutions, including the nucleonic wave functions $\phi_i$, change. For calibration purposes, these parameters are usually expressed in terms of nuclear matter properties (see Refs.~\cite{Bender2003,Reinhard2006,kortelainen2010} for definitions and discussion). In the following,  we explore two of those parameters that are common to both frameworks: $J$ - the symmetry energy, and $L$ - the symmetry energy slope for symmetric nuclear matter ~\cite{Bender2003}. That is, in this case $\alpha=\{J, L\}$,
Figure~\ref{fig:DFT Densities} shows the neutron density $\rho_n(r)$ of $^{48}$Ca for 49 different values of these parameters for both frameworks. 
The parameters are varied
within the ranges specified in Table~\ref{tab:params}, while keeping the other coupling constants fixed at their calibrated values~\cite{mcdonnell2015uncertainty, giuliani2023bayes}.

In all our case studies, the quantum many-body problem is solved using a self-consistent approach, typically involving iterative cycles in which the densities are fixed to compute the wave functions, which are then used to update the densities until convergence is reached. These calculations can become computationally intensive given their intrinsic high-dimensional structure. In the following section, we present an approach based on dimensionality reduction techniques enhanced through GP as an effective way to speed up such calculations.

\section{Dimensionality Reduction}~\label{sec: RBM}

Dimensionality reduction techniques aim to significantly reduce the active degrees of freedom when solving complex numerical problems while still retaining key features.
These approaches have been broadly applied in the physical sciences and are particularly powerful tools in the context of uncertainty quantification, where one typically needs to evaluate expensive computer models many times.
The reduced basis method (RBM)~\cite{quarteroni2015reduced,hesthaven2016certified} is such a dimensionality reduction technique that falls under the reduced order modeling umbrella~\cite{quarteroni2014reduced,brunton2022data}.

The RBM identifies a reduced set of coordinates to approximate quantities of interest and then constructs governing equations to describe how those coordinates change with respect to controlling parameters. 
These reduced coordinates are the amplitudes of a reduced basis of the solution to the parametrized equations one is solving. For the case studies we outlined in the previous section, the RBM approximation could be applied to the densities $\rho$, taking the form:
\begin{equation}\label{eq: rho expansion}
    \rho(x;\alpha)\approx \hat \rho(x;\alpha) = \sum_k^n a_k(\alpha) \rho_k(x),
\end{equation}
where $x$ denotes the set of spatial variables, usually represented by a high-dimensional grid of size $\mathcal{N}$, and $n\ll\mathcal{N}$ is a small number of coefficients $a_k(\alpha)$ that are latent reduced coordinates dependent on the control variables $\alpha$.

\begin{figure}[t!]
    \includegraphics[width=1\columnwidth]{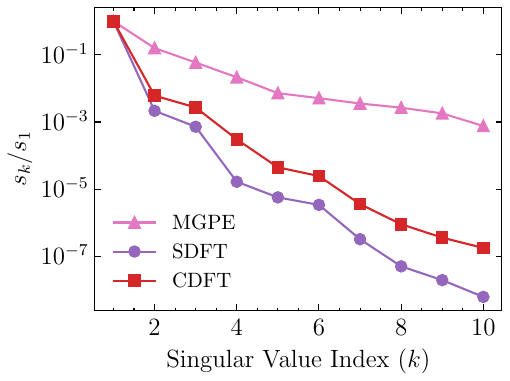}
    \caption{Decay of the singular values of the SVD of the density $\rho$ for MGPE~\eqref{eq: BGG}, and for the neutron density in $^{48}$Ca for SDFT~\eqref{eq:edf} and CDFT~\eqref{LDensity}. The singular values for each density type have been normalized with respect to the first respective value $s_1$, and their exponential decay for all three cases validates the approximate expansion in Eq.~\eqref{eq: rho expansion}. The decay is much faster in the realistic cases of SDFT and CDFT because the controlling parameters $\alpha$ have been varied over physically calibrated regions~\cite{mcdonnell2015uncertainty,giuliani2023bayes}, while the parameters for the MGPE were varied on a unphysical wide range for the sake of illustration purposes. }
    \label{fig:svdfull}
\end{figure}

The reduced basis, $\rho_k(x)$, is usually informed by previous solutions for different values of $\alpha$, called snapshots, such as those presented in Figs~\ref{fig:BGGDensities} and~\ref{fig:DFT Densities}.  To construct the basis $\rho_k$ one can apply the proper orthogonal decomposition
based on a singular value decomposition (SVD) of a set of snapshots and retaining only the $n$ singular vectors corresponding to the lowest singular values. This has the benefit of better capturing the variability across the training space with a smaller reduced basis, imposing a natural ranking of importance to the basis elements, and introducing additional numerical stability from using an orthogonal basis. A fast (exponential) decay of the  singular values will mean that a good approximation can  be obtained with a small number $n$. 

The insets of Figs.~\ref{fig:BGGDensities} and~\ref{fig:DFT Densities} show the reduced basis functions $\rho_k$ for $k=1,2,3$
for our case studies. The corresponding singular values are shown in Fig.~\ref{fig:svdfull}. For SDFT and CDFT, the decay of singular values is very fast: It is seen that three reduced basis components with the highest singular values dominate. For MGPE, the decay pattern is significantly slower. This will negatively impact the error budget of the MGPE density emulator.

Once the reduced coordinates have been identified, the next step consists of obtaining equations that describe the response of the latent variables to changes in the control variables $ a_k(\alpha)$. In many RBM applications, this is done through the Petrov-Galerkin projection of the original equations into a subspace spanned by test functions.
For a system of only one unknown function expanded by $n$ reduced basis, for example, this projection scheme would create $n$ equations to be solved for the $n$ unknown coefficients $a_k$.

This approach for constructing the reduced equations is well motivated and can provide certain accuracy guarantees, though if one wishes to apply it to the quantum many-body systems described in Sec.~\ref{sec: Quantum Systems}, it presents three main challenges:

\emph{1. Non-linear and non-affine operators}:  Here the challenge lies in the inability to precompute the Petrov-Galerkin projections in the offline (training) stage of building the emulator to avoid any scaling of size $\mathcal{N}$. This situation is encountered, e.g., for the MGPE with fractional powers of $\sigma$ \cite{bonilla2022training}.
The RBM literature explores techniques to precompute the required projections when non-affine operators are present, the Empirical Interpolation Method being one of the most widely applied
(see Ref.~\cite{rose2024} for a recent nuclear physics application). Yet when the operators are both non-linear and non-affine, 
the applicability of subspace projection-based model reduction schemes become very limited. 

\emph{2. Scalability:} The bulk of the speedup gained when applying the RBM to solve a parametrized differential equation lies in the reduction of the expensive, high-fidelity calculation to one that scales well with the size $n$ of a reduced basis.
If the exponent $\sigma$ in Eq.~\eqref{eq: BGG} is an integer instead of a fraction, the projection equations can be pre-computed to avoid any dependence on the original spatial domain of size $\mathcal{N}$.
However, for a reduced basis expanding the involved wavefunctions $\phi_i(x)$ the number of terms in the equations will scale roughly\footnote{For $n=10$ and $\sigma=4$ this accounts for exactly 715 terms on each projection equation, for example.} as $n^{\sigma}$ for a fixed $\sigma$. This, in turn, can limit the possible speed up that can be obtained by an RBM emulator \cite{giuliani2023bayes}.

\emph{3. Reduced model degrees of freedom:} Oftentimes, we are interested in understanding how the wave functions that describe the many-body system change with the controlling parameters $\alpha$. 
In many cases, however, one aims instead at estimating the global properties of the system, such as the particle density $\rho(x;\alpha)$, and a detailed information on individual states is not needed.
In such cases, if the expansion in Eq..~\eqref{eq: rho expansion} for the system density is accurate enough, we would be interested in focusing on studying only how the density coefficients $a_k(\alpha)$ of $\hat \rho(x;\alpha)$ change with the controlling parameters $\alpha$.
However, since the density is defined as a function of the solutions to the differential equation (see, for example, Eq.~\eqref{eq: density}),
it is not possible to use the Petrov-Galerkin projection scheme to separate equations for the reduced coordinates $a_k$ of $\hat \rho(x;\alpha) $ \emph{only}, and the full system has to be solved to obtain the density.
This, in turn, can limit the speed of an RBM emulator.

The problems that arise when employing the projection framework motivate the development of alternative approaches to obtain reduced equations for a given reduced-order model. The genetic programming framework explored in this Guide offers a very effective avenue. Under this framework, we build expressions that explicitly relate the small set of coefficients $a_k(\alpha)$ of the reduced basis expansion of $\rho(x;\alpha)$ in Eq.~\eqref{eq: rho expansion} to changes in the controlling parameters $\alpha$. We achieve this by solving the underlying many-body equations for a training set of parameters $\alpha$ and define the set of optimal coefficients $\{\tilde a_k(\alpha)\}$ as:
\begin{equation}\label{eq: optimal coefficients}
    \{\tilde a_k(\alpha)\} \equiv \underset{\{a_k\}}{\operatorname{argmin}} \Big|\Big|\rho(x;\alpha) - \sum_k^n a_k(\alpha) \rho_k(x)\Big|\Big|^2,
\end{equation}
that is, for a fixed basis size $n$ and each $\alpha$ in the training set, we find the coefficients $a_k$ that best approximates the respective density according to  the L2 norm. The L2 norm is defined as $||f(x)||^2 \equiv \int |f(x)|^2dx$. The GP approach is then used in a traditional machine learning regression problem: Given the observed relationship between $\tilde a_k$ and $\alpha$ in the training set, we build a functional relationship with the objective of generalizing to unobserved parameters $\alpha$. 

Through our developed framework, we are able to create very effective reduced order models by leveraging the best characteristics of both the RBM and GP methods. A good set of reduced coordinates is easily identified by the SVD traditionally performed in RBM applications, effectively compressing high-dimensional densities into a manageable set of coefficients $a_k$. The GP can then focus on discovering parsimonious expressions to describe the dynamics of this small set of coefficients as a function of $\alpha$, an easier task than if we attempted to find direct GP expressions for $\rho(x;\alpha)$. In the next section we present our results of applying this framework to the  case studies considered.

\section{Results}
\label{sec: results}

For the computational tests, we selected three basis components $n=3$ in all case studies. Both the reduced coordinates, i.e., the coefficients of the reduced basis $\rho_k(x)$, and the governing equations are learned from the training data with the parameter ranges listed in Table~\ref{tab:params}. The reduced basis $\rho_k(x)$ is obtained following a singular value decomposition on a set of self-consistent solutions, and the optimal coefficients are obtained by means of  Eq.~\eqref{eq: optimal coefficients}. We treat each coefficient as a separate function $a_k(\alpha)$ and as such we apply the LR and GP frameworks to obtain independent expression for each.

\begin{table}[htb]
\footnotesize
\caption{Summary of the GP hyper-parameters.  $P(C)$ and $P(M)$ indicate the crossover and the mutation probabilities, respectively. CF represents the continued fraction operator $d/(b+c/a)$ and RP represents the rational polynomial function $(b+d+f)/(a+c+e)$.}
\label{tab_hyperparameters_gp}  
\begin{tabularx}{\columnwidth}{lX}
\hline
\textbf{Parameter} & \textbf{Values}\\
\hline
Number of runs & 16\\
Population size & 300\\
Functions ($F$) & \{+, -, x, /, -x, CF, $x^2$, $x^{-1}$, RP, $\sqrt{x}$, $e^x$, $\log(x)$, $x^n$, $x^{1/3}$, $x^3$\}\\
Selection & Pareto tournament selection size 30\\
Genetic operators & \{subtree crossover, subtree mutation, depth-preserving subtree mutation\}\\
$P(C)$ & 0.9\\
$P(M)$ & 0.1\\
Maximum complexity & 1000 (Visitation Length \cite{visitationLength})\\
Stopping criteria & 5 minutes\\
\hline
\end{tabularx}
\end{table}
For GP we used the commercially available DataModeler system~\cite{datamodeler}. The GP parameter set used is shown in Table~\ref{tab_hyperparameters_gp},
including the hyper-parameters (HPs) used for GP, along with cross-validation settings. HPs were selected following common practice in the literature in order to avoid a computationally demanding tuning phase~\cite{winkler2025gptp21}. 
The coefficient of determination $R^2$ was used as the fitness function~\cite{c03_gp_interval_scaling_keijzer,a13_plcc_livadotiotis,c23_gptp_r2_haut,c25_assumptions_model_Kotanchek,c25_lexicase_bakurov} as it was found to converge quickly and generalize well, even when only a few data points are available.

A common practice in GP is to protect operators from undefined mathematical behavior by defining some ad hoc behavior at those points, such as, for instance, returning the value 1 in the case of a potential division by zero to make it possible for genetic programming to synthesize constants by using $x/x$~\cite{b92_gp_koza}.
However, these techniques were shown to have  several shortcomings in the vicinity of mathematical singularities~\cite{c03_gp_interval_scaling_keijzer}. In this study, programs that produced invalid values were automatically assigned  low fitness values, making them unlikely to be selected. By reducing the selection probability of solutions that produce invalid values, we expect to obtain models whose fitness landscape is less ``sharp''~\cite{c25_sharpness_bakurov}.

In addition to the more expressive LR models obtained via the transformation pipeline described in Sec.~\ref{method_lr}, we also include a standard second-order polynomial model composed by squared terms and multiplicative interactions between input features. This model represents the simplest form of non-linear regression and serves as a baseline for comparison. 
For example, modeling the coefficients of the neutron density $a_k(\alpha)$ as a function of parameters of $J$ and $L$, the second order polynomial is:
\begin{equation}\label{eq: poly-2}
\beta_{0} + \beta_{1}\,L + \beta_{2}\,J + \beta_{3}\,L^{2} + \beta_{4}\,L\,J + \beta_{5}\,J^{2},
\end{equation}
where the coefficients $\beta_i$ are adjustable parameters.

\begin{table}[h]
\caption{Summary of the four methods used to calculate the densities of the case studies.
The Full Order Model (FOM) represents the high-fidelity MGPE and DFT calculations. Poly-2 is the second-order expansion in model parameters $\alpha$.
}
\label{tab:methods}
\footnotesize
\begin{tabularx}{\columnwidth}{@{}>{\arraybackslash}m{0.27\columnwidth}|>{\arraybackslash}X@{}}
\hline
\parbox[t]{\linewidth}{\textbf{Method}} & \textbf{Specification}\\
\hline
\\[-5pt]
\parbox[t]{\linewidth}{FOM}
  & Finite element method with an iterative solver with grid size of 300. \\\\[-5pt]
\parbox[t]{\linewidth}{LR}
  & Regularized LR with custom feature transformation and selection pipeline, built using scikit-learn API~\cite{a11_scikit_pedregosa,c13_sklearn_api_buitinck}.\\\\[-5pt]
\parbox[t]{\linewidth}{Poly-2}
  & Second order polynomial fit without regularization built using scikit-learn API~\cite{a11_scikit_pedregosa,c13_sklearn_api_buitinck}. \\\\[-5pt]
\parbox[t]{\linewidth}{GP}
  & Symbolic expressions generated using the commercially available DataModeler system~\cite{datamodeler}.\\
\hline
\end{tabularx}
\end{table}

Once the reduced models were trained (see Appendix~\ref{ssec:equations} for the obtained mathematical expressions, and Table~\ref{tab:methods} for the model descriptions), we tested their generalization ability on a test set of previously unexplored combinations of parameters $\alpha$. The first test consisted of qualitatively comparing the predictions of each method with the optimal coefficients defined in Eq.~\eqref{eq: optimal coefficients}. 

\begin{figure}[htb]
    \includegraphics[width=1\columnwidth]{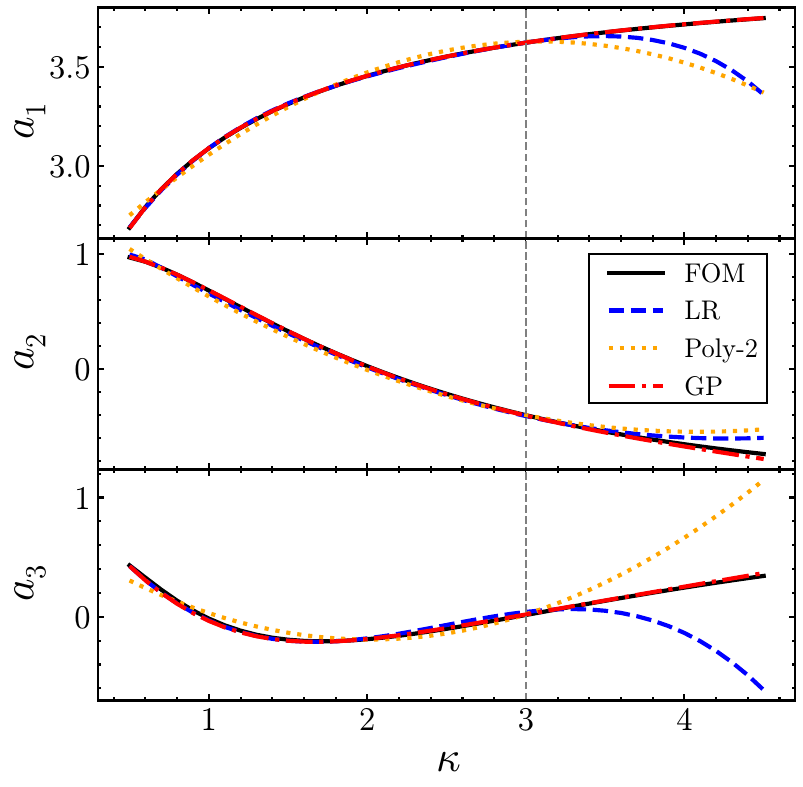}
    \caption{Evolution of the three coefficients $a_k$ in the density reduced order expansion~\eqref{eq: rho expansion} for the MGPE~\eqref{eq: BGG} as a function of the dimensionless parameter $\kappa$ at $q=1$ and $\sigma=1.25$. The FOM  results 
    (solid line), are compared to the LR (dashed line), Poly-2 (dotted line), and GP (dash-dotted) calculations. The vertical dashed  line at $\kappa=3$ divides the region into interpolation ($\kappa\leq 3$) and extrapolation ($\kappa>3$), with GP being the only method that fully recovers the behavior in the extrapolated region. We obtained similar results when varying $q$ and $\sigma$ while keeping  other parameters constant. }
    \label{fig:bgg_extrapolation}
\end{figure}
In the case of the modified MGPE~\eqref{eq: BGG}, Fig.~\ref{fig:bgg_extrapolation} shows the dependency of the three basis coefficients $a_k$ as a function of $\kappa$ when the other two parameters $\{q,\sigma \}$ are kept constant. 
As can be seen, GP has appreciably better performance in reproducing these optimal coefficients compared to the other methods. In particular, it excels in extrapolating beyond the dashed vertical line which depicts the maximum parameter value $\kappa=3$ used for the training of models, see Table~\ref{tab:params}. This suggests that GP has learned expressions that can extrapolate the reduced dynamics with less overfitting. 

\begin{figure}[htb]
    \includegraphics[width=1\linewidth]{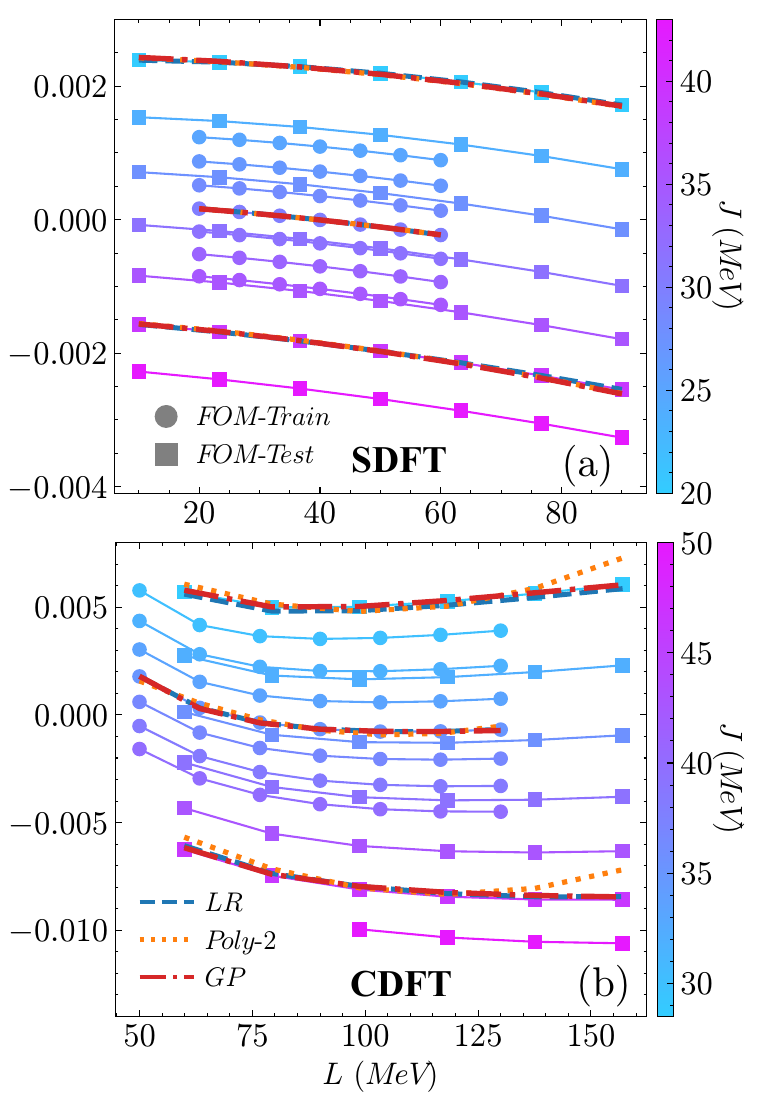}
    \caption{Evolution of the third coefficient $a_3(L,J)$ in the $^{48}$Ca neutron density reduced basis expansion~\eqref{eq: rho expansion} as a function of the two controlling parameters $L$ and $J$ for the case of SDFT (a) and CDFT (b). The color grading curves represent the optimal coefficient $a_3$~\eqref{eq: optimal coefficients} obtained from the FOM calculations, with circles marking training parameters  and squares marking the testing parameters. The dashed and dotted lines show the LR, Poly-2, and GP results for three  values of $J$, one in the training region and two in the testing (extrapolation) region.}
    \label{fig:CoeffsSyrmeRMF}
\end{figure}

\begin{figure}[htb!]
 \includegraphics[width=0.9\linewidth]{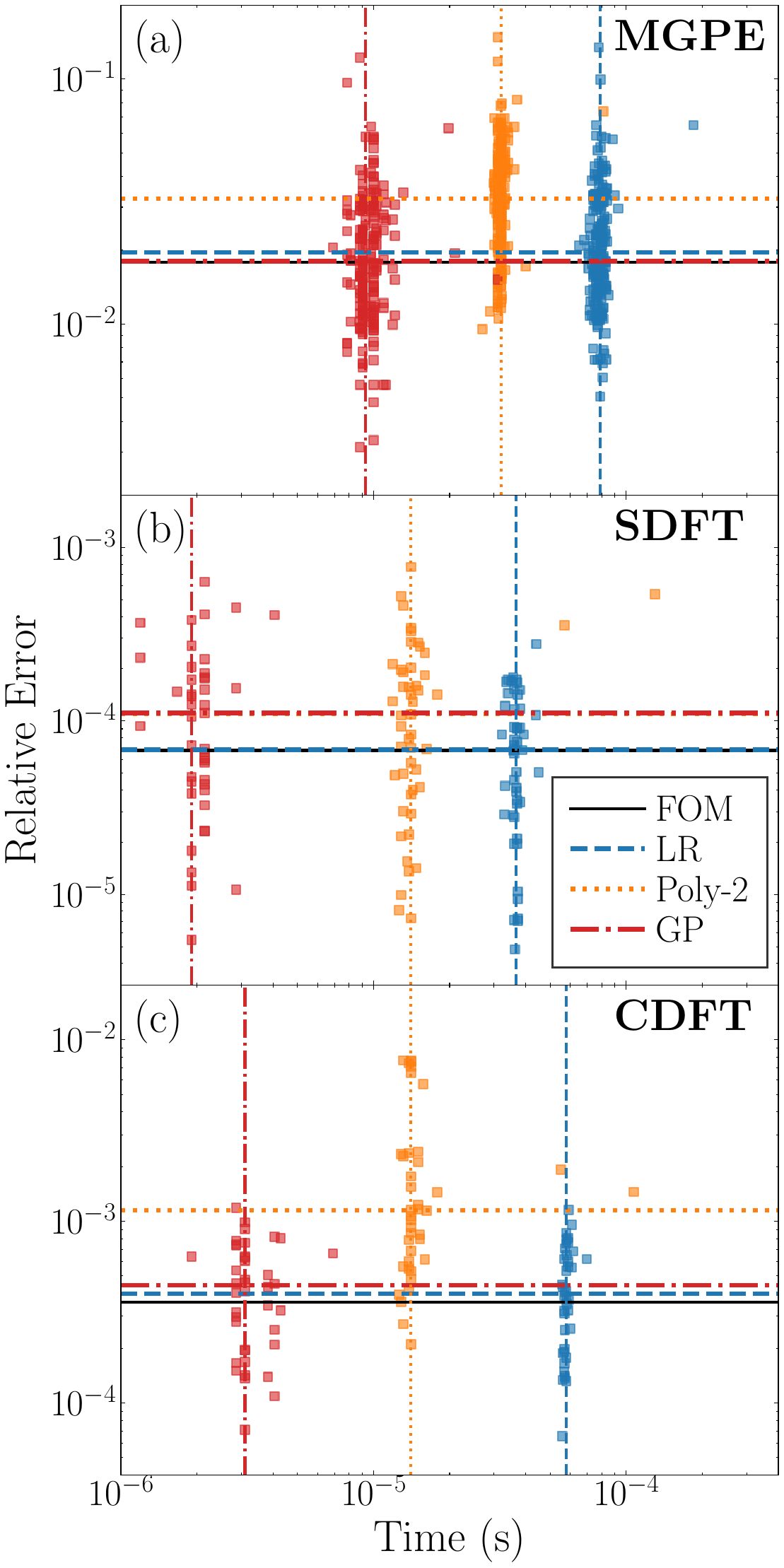}
    \caption{\label{fig: Global CAT}
   Computational Accuracy vs Time plot for (a) MGPE,
   (b) SDFT, and (c) CDFT density calculations.  The squares show individual calculations for the testing parameters using the reduced order models with  LR, Poly-2, and GP. A total of 212 (a), 49 (b), and 40 (c) testing parameters are shown. The horizontal lines represent the median relative error \eqref{eq: relative error}  for each method, with the FOM line representing the ``best case scenario". The vertical lines represent the average time of the calculations for each of the methods. For FOM, the time is  orders of magnitude slower and hence it is not shown.
    }
\end{figure}

For the $^{48}$Ca neutron density in SDFT and CDFT, 
Fig.~\ref{fig:CoeffsSyrmeRMF} shows the evolution of the third coefficient $a_3(L, J)$ in the reduced basis expansion~\eqref{eq: rho expansion} as a function of the two controlling parameters $L$ and $J$. The figure shows the optimal (target) coefficients, obtained from the FOM calculations~\eqref{eq: optimal coefficients} as gradient-colored solid lines, 
and compares three of them with the calculations using the different methods listed in Table~\ref{tab:methods}, as colored dashed lines. 
In SDFT, all models demonstrate high generalization performance in predicting the coefficient $a_3(L, J)$. In contrast, for CDFT, the second-degree polynomial model does not extrapolate accurately, particularly at higher values of $L$ ($L > 125$).
Although visual comparison may suggest comparable accuracy between custom LR and GP, the symbolic expressions in Appendix~\ref{ssec:equations} reveal that GP-derived models are significantly more compact. This contributes not only to improved inference efficiency but also to greater interpretability, reinforcing GP's advantage in scientific model discovery.

The second test we developed for the methods specified in Table~\ref{tab:methods} focused on a quantitative assessment of their performance. For our purpose, the two main qualities we seek in a reduced-order model are its speed and its accuracy. We define the speed of the emulator -- once it has been trained and all relevant quantities have been precomputed -- as the time it takes to map controlling parameters $\alpha$ to the quantities of interest, in this case the amplitudes of the reduced basis for the particle densities for the two quantum many-body systems. We quantify the accuracy for a given parameter $\alpha_i$ through the relative root mean squared error (or $L^2$ norm) between the density obtained from the high-fidelity solver and the one obtained from the reduced order model:
\begin{equation}\label{eq: relative error}
   \text{Relative Error}(\alpha_i) = \frac{\big|\big|\rho(x;\alpha_i)-\hat\rho(x;\alpha_i)\big|\big|}{\big|\big|\rho(x;\alpha_i)\big|\big|},
\end{equation}
where $\hat \rho$ is the approximation in Eq.~\eqref{eq: rho expansion} with coefficients $a_k(\alpha_i)$ predicted by the respective method.

Figure~\ref{fig: Global CAT} shows the trade-off between accuracy and speed, quantified as the relative error (vertical axis) and the inference times (horizontal axis) for the test parameters across the studied systems. 
Individual predictions for the testing parameters are represented by squares.
The plots clearly show that the GP-based models achieve the fastest inference times while maintaining a low error. In the MGPE case, GP outperforms all other models in both accuracy and speed. Both GP and custom LR achieve relative errors that are at the theoretical limit of $\sim 1.5\%$ (horizontal black line), which means that any further improvement in precision will come from increasing the number of bases beyond $n=3$, and not from obtaining more accurate coefficients $a_k(\alpha)$. 

For the SDFT framework,  LR yields a smaller error -- on top of the theoretical limit of $\sim 0.007\%$ -- but at the cost of significantly longer inference times; GP, however, performs inference in more than one order of magnitude faster while only slightly compromising the prediction quality with an error of $\sim 0.01\%$. In the CDFT case, GP and custom LR report similarly small errors -- also comparable with the theoretical limit of $\sim0.03\%$ -- but GP does so with more than an order of magnitude faster inference than LR. In all cases, all methods provide appreciable speedups between 4 and 6 orders of magnitude when compared with the standard FOM solvers.

It is worth noting that the faster inference times of GP-based models are a direct consequence of their compact functional form (see Appendix~\ref{ssec:equations}) and reduced data preprocessing requirements. In particular, both custom LR and Poly-2 models require 0-1 scaling of the input data at each inference event, which contributes to higher run-times. On the other hand, GP models operate directly on raw input values. This justifies the fact that even a relatively simple second-order polynomial lags behind GP in terms of run-times.

\section{Conclusions and Outlook} \label{sec: conclusions}
In this Guide we developed and presented a general framework based on Genetic Programming to aid in the construction of reduced order models used in expensive nuclear physics computations. The GP approach employs evolutionary algorithms to propose, test and evolve a set of candidate governing equations that control how response variables change with respect to controlling variables. With the specific application to three many-body case studies, we used the GP approach to create the mathematical expressions that describe the change in the reduced coordinates of high-dimensional solutions. By using the GP approach to build the equations, we have been able to successfully craft effective reduced-order models for these systems, allowing speedup gains of 5 to 6 orders of magnitude in comparison to their respective full-order solvers at a negligible loss in accuracy. 

It should be noted that although surrogate models based on reduced-order modeling, such as the reduced basis method used here, have provided spectacular success in terms of many orders of magnitude computational speedup gains at negligible loss in accuracy, certain classes of problems are not amenable to these approaches. Furthermore, in general, the development of the surrogate model requires an appreciable upfront effort to construct, often necessitating access to and manipulation of the underlying complex equations of the full system. 

Future work on using information on the underlying operators to inform the evolution algorithm and the construction of implicit equations~\cite{wang2015multi,izzo2017differentiable,schmidt2009symbolic} relating the control variables and the response variables could further improve the performance of the built surrogate models and also help in the interpretability of the obtained equations obtained through GP. The development of a pipeline to straightforwardly build effective and more interpretable reduced order models for nuclear science will enable a broad adoption of uncertainty quantification approaches~\cite{phillips2021get}, and allow the theory to provide data across thousands of nuclei necessary to face modern challenges in this new era of discovery~\cite{LRP2023}.

\section*{Acknowledgements}

We are grateful to Edgard Bonilla for useful discussions.
This work was supported by the U.S.
Department of Energy, Office of Science and Office of Nuclear
Physics under Awards Nos. DOE-DE-SC0013365, SC0024586,
and DE-SC0023175 (Office of Advanced Scientific Computing Research and Office of Nuclear Physics, Scientific Discovery through Advanced Computing). W.B. gratefully acknowledges funding provided through the John R. Koza endowment to Michigan State University.

\section*{Data availability}
All data that support the findings of this study are included within the article and are openly available \cite{github}.


\appendix
\section{Appendix}\label{sec:appendix}
In this work, we use genetic programming (GP) and compare its performance to construct nonlinear and interpretable emulators against an augmented version of linear regression (LR).
LR is one of the most popular and frequently used SML tools for regression. It is known for its simplicity, computational efficiency and interpretability. It assumes a linear relationship between $k$ input variables/features ($x$) and the output target ($y$). Formally:
$y = \beta_0 + \beta_1 x_1 + \beta_2 x_2 + \ldots + \beta_n x_n$, where $y$ is the target, $\beta_0$ is the intercept, and $\beta_k$ is a learnable weight associated with input feature $x_k$. The learnable weights and the intercept are usually estimated from the data using Ordinary Least Squares (OLS) estimation procedure to minimize the sum of squared residuals (i.e., difference between the observed target values in the training dataset, and the values predicted by the linear model). Formally: $\min_{\beta} \sum_{i=1}^{n} (y_i - \hat{y}_i)^2$, where $\hat{y}_i$ is the model's prediction for data instance $i$.

The assumption of linearity reduces the ability of LR
to produce models in complex and multi-dimensional feature spaces.
This structural rigidity makes it fundamentally different from symbolic regression methods, like GP, which simultaneously search for both the functional form and parameters of a mathematical expression to best fit the data.
To overcome this limitation and enhance the expressiveness of LR, we use a custom feature transformation pipeline. This pipeline integrates tailored feature transformations, including several mathematical functions (such as logarithmic and exponential functions, roots, powers, etc.), and feature interactions (products and quotients). These transformed features are concatenated with the original inputs and the bias term. 
However, when the feature space is inflated and comprises potentially highly non-linear relationships, the risk of overfitting becomes particularly high. For this reason, we include several levels of feature selection and weight-decay regularization in our pipeline. 
In this way, our goal is to make LR more competitive (and comparable) to GP in the capture of complex patterns. Additionally, we include a standard second-order polynomial linear regression model with multiplicative interaction terms as a baseline for comparison. 

We use randomized cross-validation to find the most suitable hyper-parameters for our pipeline. This includes the choice of feature transformations, feature interactions, feature selection and regression model-specific parameters such as regularization strength (these aspects are described in detail in the following sections). In total, 250 parameter settings are sampled and the best models are selected after aggregating validation errors from 3-fold cross validation. 
To perform the experiments, we use predictive models, cross-validation and pipeline construction tools provided by scikit-learn~\cite{a11_scikit_pedregosa,c13_sklearn_api_buitinck}. Full details about our LR models, pipeline composition and training are provided in our source code available at \cite{github}.

\subsection{Feature Transformation}\label{ssec:feature_transformation}
Different combination of feature transformations for LR are seen as different hyper-parameter values in our pipeline. These transformations range from polynomials, such as $\{x^2, x^3\}$, to more complex terms $\{\sqrt{x}, \sqrt[3]x, e^x, ln(x), x^{-1}\}$. We include two types of feature interactions, one through multiplication (eg. $x_1\times x_2$) and another through division (eg. $x_1/x_2$). In such a way, LR can be equipped with simpler or more complex feature spaces, balancing model interpretability and complexity. 
Exploring the space of feature transformations in this manner allows for a fairer comparison with GP. Recall that GP operates on a highly expressive feature space, and can employ arbitrary powers (e.g., square and cube), roots (e.g., square and cube roots), logarithms, and exponentials. In this sense, our custom feature engineering process for LR allows it to approximate the expressiveness of GP, ensuring that the comparison between methods is equitable.

\subsection{Feature Selection}\label{ssec:feature_selection}
We are aware that inflating the input feature space by applying a variety of transformations introduces significant challenges in the SML context. Deliberately increasing the proportion of features relative to the number of data instances (which remains fixed) is expected to penalize the generalization of estimated regression coefficients (i.e., leads to overfitting). 
To mitigate this challenge, we incorporate up to two consecutive steps of feature selection.
In the first step, we allow the pipeline to perform feature selection that uses a tree-based ensemble strategy, called Extremely Randomized Trees (ExtraTrees)~\cite{a06_extra_trees_geurts}, known for its robustness and computational efficiency. The ExtraTrees method fits multiple randomized decision trees to the training data in parallel, and then aggregates their results. 
The inference of randomized trees relies on randomly choosing a set of splitting points and then selects the one that minimizes a given loss function, instead of exhaustively enumerating all possible splits.
Among them is the feature importance ranking provided by each tree. Note that in decision tree learning, features that are more frequently used at the top of trees (where the first splits occur) are considered more important. 
The aggregation across multiple trees allows to reduce variance and dependency from a single model's output~\cite{a00_bias_variance_decomposition_domingos}.
The second step of feature selection is optional and implements a refinement of the previous step. It involves a linear regression model to make the selection based on the predictive power of the remaining features and their linear relationship with the target variable.
This results in potentially simpler and possibly more interpretable models.

\subsection{Regularization}\label{ssec:appendix_ft}
While feature selection helps to reduce the dimensionality of the input space, it does not fully eliminate the risk of overfitting. To further improve the generalization ability of our models, we also incorporate regularization techniques. Regularization adds an explicit penalty term to the regression loss function, discouraging overly complex models and helping to prevent the model from fitting the noise. Several levels of regularization are included,  L1 regularization (Lasso)~\cite{a96_lasso_tibshirani}, L2 regularization (Ridge)~\cite{a70_ridge_hoer}, and a combination of the two (ElasticNet)~\cite{a05_elasticnet_zou}. 
L1 regularization promotes model sparsity and is useful in high-dimensional tasks (i.e. when there are many input features), especially if many are expected to be irrelevant. L2 regularization is frequently used when the input features are expected to be informative but their impact has to be reduced to avoid overfitting, and/or if they are highly correlated (multicollinear). Formally, a L1 term is given by $\lambda_1\sum_{k=0}^{n} |\beta_k|$, while a L2 term is given by $\lambda_2\sum_{k=0}^{n} \beta_k^2$.
ElasticNet uses combines and balances both L1 and L2.

Note that regularized linear regression has long been used in scientific applications. Sparse-identification of non-linear dynamics (SINDy) is a popular method for discovering governing equations in dynamic systems~\cite{brunton2016discovering}. Essentially, it utilizes linear regression with the L1 regularizing term to promote sparsity, yielding smaller models, which can offer better interpretability,  considered an appealing property. SINDy has been successfully applied to find governing equations in many different applications~\cite{hoffmann2019reactive,tran2017exact,fasel2022ensemble,jiang2021modeling,shea2021sindy}.
Nevertheless, SINDy-like approaches present an obvious disadvantage: nonlinear relationships are combinatorial in input variables and must be manually pre-defined, requiring some degree of expertise. This can be particularly limiting when applied to problems where valuable nonlinear relationships are unexpected and is one of the advantages we expect from GP.

\clearpage
\onecolumngrid

\subsection{Expressions}
\label{ssec:equations}

In this Appendix we present the discovered mathematical expressions for the case studies described in Sec.~\ref{sec: Quantum Systems} when using the three surrogate methods (LR, Poly-2, and GP) summarized in Table~\ref{tab:methods}. The coefficients $a_k$ are dimensionless in all cases since the respective dimensions of the densities in Eq.~\eqref{eq: rho expansion} is carried out by the respective principal components $\rho_k$. The parameters of the MGPE are themselves dimensionless, so we write direct equations between them and the response variables $a_k$. For the case of SDFT and CDFT we write the equations in terms of the dimensionless parameters $\tilde J = J/\text{MeV}$ and $\tilde L = L/\text{MeV}$. 

\subsubsection{MGPE}
\label{ssec:eq_gross_pitaevskii}


\begin{align*}
a_1(\kappa , q, \sigma)_{LR} =\ &- 1.383 \kappa^{4} - 0.099 \kappa^{3} q + 0.034 \kappa^{3} \sigma + 4.001 \kappa^{3} + 0.057 \kappa^{2} q^{2} + 0.311 \kappa^{2} q \sigma - 0.088 \kappa^{2} q \\
&+ 0.015 \kappa^{2} \sigma^{2} - 0.246 \kappa^{2} \sigma - 4.485 \kappa^{2} - 0.015 \kappa q^{3} + 0.110 \kappa q^{2} \sigma - 0.143 \kappa q^{2} + 0.118 \kappa q \sigma^{2} \\
&- 0.873 \kappa q \sigma + 0.619 \kappa q - 0.040 \kappa \sigma^{3} - 0.012 \kappa \sigma^{2} + 0.444 \kappa \sigma + 2.587 \kappa \\
&- 0.018 q^{4} - 0.037 q^{3} \sigma + 0.068 q^{3} - 0.061 q^{2} \sigma^{2} + 0.066 q^{2} \sigma - 1.148 \cdot 10^{-3} q^{2} \\
&- 0.452 q \sigma^{3} + 0.745 q \sigma^{2} + 0.129 q \sigma - 0.512 q + 9.650 \cdot 10^{-4} \sigma^{4} + 0.254 \sigma^{3} - 0.399 \sigma^{2} - 0.067 \sigma \\
&+ 3.030
\end{align*}

\begin{align*}
a_2(\kappa , q, \sigma)_{LR} =\ &+ 0.002 \sqrt{\kappa} \sqrt{q} + 0.032 \sqrt{\kappa} q^{2} + 0.469 \sqrt{\kappa} q - 0.107 \sqrt{\kappa} - 0.014 \kappa^{2} q^{2} - 0.083 \kappa^{2} q \\
&+ \frac{0.324 \kappa^{2}}{(\sqrt{q} + 1)^{0.5}} + \frac{0.526 \kappa^{2}}{(q + 1)^{0.5}} + \frac{0.505 \kappa^{2}}{(q^{2} + 1)^{0.5}} - 0.210 \kappa \sqrt{q} - 0.051 \kappa \sqrt{\sigma} \\
&- 0.003 \kappa \sigma - \frac{0.383 \kappa}{(\sqrt{q} + 1)^{0.5}} - \frac{0.723 \kappa}{(q + 1)^{0.5}} - \frac{0.356 \kappa}{(q^{2} + 1)^{0.5}} - \frac{0.037 \kappa}{(\sqrt{\sigma} + 1)^{0.5}} \\
&- \frac{0.065 \kappa}{(\sigma + 1)^{0.5}} - \frac{4.500 \cdot 10^{-3} \kappa}{(\sigma^{2} + 1)^{0.5}} - 1.300 \kappa - 0.010 \sqrt{q} \sigma^{2} - 0.011 \sqrt{q} \sigma \\
&+ \frac{0.040 \sqrt{q}}{(\kappa^{2} + 1)^{0.5}} + \frac{0.079 \sqrt{q}}{(\sigma^{2} + 1)^{0.5}} - \frac{0.065 q^{2}}{(\sqrt{\kappa} + 1)^{0.5}} - 0.176 q \sigma^{2} - 0.005 q \sigma \\
&+ \frac{5.800 \cdot 10^{-3} q}{(\kappa^{2} + 1)^{0.5}} + \frac{0.160 q}{(\sigma^{2} + 1)^{0.5}} + \frac{0.033 \sigma^{2}}{(\kappa + 1)^{0.5}} + \frac{0.078 \sigma^{2}}{(q + 1)^{0.5}} \\
&+ \frac{0.022 \sigma^{2}}{(q^{2} + 1)^{0.5}} + \frac{0.063 \sigma}{(\kappa + 1)^{0.5}} + 0.802
\end{align*}

\begin{align*}
a_3(\kappa , q, \sigma)_{LR} =\ &-2.819\kappa^{3} + 0.358\kappa^{2}q + 0.196\kappa^{2}\sigma + 5.669\kappa^{2} + 0.269\kappa q^{2} + 0.835\kappa q\sigma \\
&- 1.839\kappa q + 1.482\cdot 10^{-3} \kappa \sigma^{2} - 0.654\kappa \sigma - 2.296\kappa - 0.027q^{3} \\
&- 0.090q^{2} \sigma + 0.036q^{2} + 0.245q \sigma^{2} - 0.662q \sigma + 0.677q \\
&+ 0.033\sigma^{3} - 0.175\sigma^{2} + 0.421\sigma - 0.059
\end{align*}


\begin{align*}
a_1(\kappa, q, \sigma)_{Poly-2} =\ 
&- 1.023\,\kappa^{2} + 0.154\,\kappa q + 0.006\,\kappa \sigma + 1.778\,\kappa + 0.016\,q^{2} + 0.210\,q \sigma \\
&- 0.390\,q - 3.012 \cdot 10^{-3}\,\sigma^{2} - 0.104\,\sigma + 3.020
\end{align*}

\begin{align*}
a_2(\kappa, q, \sigma)_{Poly-2} =\ 
&0.971\,\kappa^{2} + 0.204\,\kappa q - 0.033\,\kappa \sigma - 2.629\,\kappa - 0.072\,q^{2} - 0.316\,q \sigma \\
&+ 0.484\,q + 0.015\,\sigma^{2} + 0.171\,\sigma + 0.740
\end{align*}

\begin{align*}
a_3(\kappa, q, \sigma)_{Poly-2} =\ 
&1.695\,\kappa^{2} - 0.806\,\kappa q - 0.041\,\kappa \sigma - 1.216\,\kappa + 0.086\,q^{2} - 0.101\,q \sigma \\
&+ 0.316\,q - 1.167 \cdot 10^{-2}\,\sigma^{2} + 0.070\,\sigma + 0.021
\end{align*}

\begin{align*}
a_1(\kappa , q, \sigma)_{GP} =\ & 4.162 - \frac{0.012}{\kappa} - 0.015k - \frac{1.676}{0.738 + \kappa} + 0.007q + \frac{0.028q}{0.264 + \kappa} \\
& - \frac{0.230q}{\kappa + 0.142q + \frac{1}{2\sigma} + \sigma} - \frac{0.372q}{3 + \kappa + \sigma^{2}} + \frac{0.093q}{\sigma + \frac{12.828 + q}{10.738 + \sigma}}
\end{align*}

\begin{align*}
a_2(\kappa , q, \sigma)_{GP} =\ & -1.051 - 1.973 \cdot 10^{-4} \, \kappa^{4} + \frac{6.549}{3 + k^{2}} - 0.016kq + 0.002kq^{2} \\
& + 0.031q\sigma^{1/3} - 0.007q\sigma - \frac{0.043q}{\kappa + \kappa^{2}\sigma} - \frac{0.163k^{3}q^{2}\sigma}{(4\kappa^{2} + \sigma^{2})^{2}} \\
& + \frac{0.559kq}{\frac{1}{\kappa} + \kappa^{2} + \frac{q}{1 + \kappa} + \frac{1}{\sigma} + \sigma^{2}}
\end{align*}

\begin{align*}
a_3(\kappa , q, \sigma)_{GP} =\ & 1.276 - 0.079k + 0.020k^{2} - \frac{4.077}{\frac{2.036}{\kappa} + \kappa} - 0.199q - 0.019kq \\
& + 0.133k^{\frac{q}{\left(\frac{1}{\kappa} + \kappa\right)(\kappa + \sigma)}}q + 1.725 \cdot 10^{-4} \kappa^{2}q\sigma^{3} + \frac{0.573q}{\frac{1}{\kappa} + \kappa^{3} + \sigma}
\end{align*}

\subsubsection{SDFT}
\label{ssec:eq_skryme_dft}

\begin{align*}
a_1({\tilde L}, {\tilde J})_{LR} =\ & -1.123 \cdot 10^{-4} {\tilde L}^3 {\tilde J}^3 + 2.206 \cdot 10^{-4} {\tilde L}^3 {\tilde J}^2 - 8.261 \cdot 10^{-5} {\tilde L}^3 {\tilde J} - 8.083 \cdot 10^{-5} {\tilde L}^3 \\
& + 2.071 \cdot 10^{-4} {\tilde L}^2 {\tilde J}^3 - 3.914 \cdot 10^{-4} {\tilde L}^2 {\tilde J}^2 + 1.013 \cdot 10^{-4} {\tilde L}^2 {\tilde J} + 6.635 \cdot 10^{-4} {\tilde L}^2 \\
& - 4.132 \cdot 10^{-5} {\tilde L} {\tilde J}^3 - 1.743 \cdot 10^{-4} {\tilde L} {\tilde J}^2 + 0.00199 {\tilde L} {\tilde J} - 0.0133 {\tilde L} \\
& - 6.594 \cdot 10^{-5} {\tilde J}^3 + 2.775 \cdot 10^{-4} {\tilde J}^2 - 0.005 {\tilde J} + 1.007
\end{align*}

\begin{align*}
a_2({\tilde L}, {\tilde J})_{LR} =\ & -5.273 \cdot 10^{-5} {\tilde L}^{3} {\tilde J}^{3} + 1.044 \cdot 10^{-4} {\tilde L}^{3} {\tilde J}^{2} - 3.772 \cdot 10^{-5} {\tilde L}^{3} {\tilde J} - 4.313 \cdot 10^{-5} {\tilde L}^{3} \\
&+ 8.843 \cdot 10^{-5} {\tilde L}^{2} {\tilde J}^{3} - 1.693 \cdot 10^{-4} {\tilde L}^{2} {\tilde J}^{2} + 2.005 \cdot 10^{-5} {\tilde L}^{2} {\tilde J} + 4.320 \cdot 10^{-4} {\tilde L}^{2} \\
&- 4.879 \cdot 10^{-6} {\tilde L} {\tilde J}^{3} - 1.558 \cdot 10^{-4} {\tilde L} {\tilde J}^{2} + 1.127 \cdot 10^{-3} {\tilde L} {\tilde J} - 7.211 \cdot 10^{-3} {\tilde L} \\
&- 2.011 \cdot 10^{-5} {\tilde J}^{3} + 1.634 \cdot 10^{-4} {\tilde J}^{2} + 5.522 \cdot 10^{-4} {\tilde J} + 2.880 \cdot 10^{-3}
\end{align*}

\begin{align*}
a_3({\tilde L}, {\tilde J})_{LR} =\ & -1.171 \cdot 10^{-6} {\tilde L}^{3} {\tilde J}^{3} + 1.864 \cdot 10^{-6} {\tilde L}^{3} {\tilde J}^{2} - 2.853 \cdot 10^{-6} {\tilde L}^{3} {\tilde J} + 7.135 \cdot 10^{-6} {\tilde L}^{3} \\
&+ 1.081 \cdot 10^{-5} {\tilde L}^{2} {\tilde J}^{3} - 1.948 \cdot 10^{-5} {\tilde L}^{2} {\tilde J}^{2} + 3.628 \cdot 10^{-5} {\tilde L}^{2} {\tilde J} - 1.362 \cdot 10^{-4} {\tilde L}^{2} \\
&- 1.723 \cdot 10^{-5} {\tilde L} {\tilde J}^{3} + 6.430 \cdot 10^{-5} {\tilde L} {\tilde J}^{2} - 1.558 \cdot 10^{-4} {\tilde L} {\tilde J} - 2.160 \cdot 10^{-4} {\tilde L} \\
&- 7.182 \cdot 10^{-6} {\tilde J}^{3} + 1.261 \cdot 10^{-4} {\tilde J}^{2} - 2.203 \cdot 10^{-3} {\tilde J} + 1.237 \cdot 10^{-3}
\end{align*}


\begin{align*}
a_1({\tilde L}, {\tilde J})_{\text{Poly-2}} =\ & 5.200 \cdot 10^{-4} {\tilde L}^2 + 1.710 \cdot 10^{-3} {\tilde L} {\tilde J} - 0.013 {\tilde L} + 4.656 \cdot 10^{-5} {\tilde J}^2 - 0.005 {\tilde J} + 1.007
\end{align*}

\begin{align*}
a_2({\tilde L}, {\tilde J})_{\text{Poly-2}} =\ & 3.470 \cdot 10^{-4} {\tilde L}^2 + 9.170 \cdot 10^{-4} {\tilde L} {\tilde J} - 0.007 {\tilde L} + 4.583 \cdot 10^{-5} {\tilde J}^2 + 1.000 \cdot 10^{-3} {\tilde J} + 3.000 \cdot 10^{-3}
\end{align*}

\begin{align*}
a_3({\tilde L}, {\tilde J})_{\text{Poly-2}} =\ & -1.130 \cdot 10^{-4} {\tilde L}^2 - 8.275 \cdot 10^{-5} {\tilde L} {\tilde J} - 2.380 \cdot 10^{-4} {\tilde L} + 1.330 \cdot 10^{-4} {\tilde J}^2 - 0.002 {\tilde J} + 1.000 \cdot 10^{-3}
\end{align*}


\begin{align*}
a_1({\tilde L}, {\tilde J})_{GP} =\ 
& 0.467 + \frac{697.688}{1243.500 + {\tilde L}} - 5.997 \cdot 10^{-4} {\tilde J} + 4.271 \cdot 10^{-6} {\tilde L} {\tilde J}
\end{align*}

\begin{align*}
a_2({\tilde L}, {\tilde J})_{GP} =\ 
& -0.261 + \frac{291.699}{1092.852 + {\tilde L}} + 2.420 \cdot 10^{-5} {\tilde J} + 2.298 \cdot 10^{-6} {\tilde L} {\tilde J}
\end{align*}

\begin{align*}
a_3({\tilde L}, {\tilde J})_{GP} = -0.032 + 2.043 \cdot 10^{-6} {\tilde L} - 7.054 \cdot 10^{-8} {\tilde L}^2 - 2.066 \cdot 10^{-7} {\tilde L} {\tilde J} + \frac{4.975}{126 + {\tilde J}}
\end{align*}

\subsubsection{CDFT}\label{ssec:eq_rmf_dft}

\begin{align*}
a_1({\tilde L}, {\tilde J})_{LR} =\ &- 0.004 \sqrt{{\tilde L}} \sqrt{{\tilde J}} - 0.004 \sqrt{{\tilde L}} {\tilde J}^2 + 0.035 \sqrt{{\tilde L}} {\tilde J} - 0.079 \sqrt{{\tilde L}} + 1.939 \cdot 10^{-4} {\tilde L}^2 \sqrt{{\tilde J}} \\
&- 0.003 {\tilde L}^2 {\tilde J}^2 + 0.009 {\tilde L}^2 {\tilde J} - 0.005 {\tilde L}^2 + 0.002 {\tilde L} \sqrt{{\tilde J}} + 0.008 {\tilde L} {\tilde J}^2 - 0.037 {\tilde L} {\tilde J} + 0.036 {\tilde L} \\
&+ 3.272 \cdot 10^{-4} \sqrt{{\tilde J}} \ln({\tilde L}) + 0.001 \sqrt{{\tilde J}} - 7.290 \cdot 10^{-5} {\tilde J}^2 \ln({\tilde L}) - 0.001 {\tilde J}^2 \\
&- 0.001 {\tilde J} \ln({\tilde L}) - 0.008 {\tilde J} + 0.003 \ln({\tilde L}) + 3.497 \cdot 10^{-5} \ln({\tilde J}) + 1.042
\end{align*}

\begin{align*}
a_2({\tilde L}, {\tilde J})_{LR} =\ & - 7.494 \cdot 10^{-4} \sqrt{{\tilde L}} \sqrt{{\tilde J}} - 5.630 \cdot 10^{-3} \sqrt{{\tilde L}} {\tilde J}^2 + 0.025 \sqrt{{\tilde L}} {\tilde J} - 0.042 \sqrt{{\tilde L}} \\
& + 3.732 \cdot 10^{-4} {\tilde L}^2 \sqrt{{\tilde J}} - 1.653 \cdot 10^{-3} {\tilde L}^2 {\tilde J}^2 + 4.115 \cdot 10^{-3} {\tilde L}^2 {\tilde J} - 7.376 \cdot 10^{-4} {\tilde L}^2 \\
& - 4.778 \cdot 10^{-4} {\tilde L} \sqrt{{\tilde J}} + 6.697 \cdot 10^{-3} {\tilde L} {\tilde J}^2 - 0.022 {\tilde L} {\tilde J} + 0.012 {\tilde L} \\
& + 9.966 \cdot 10^{-5} \sqrt{{\tilde J}} \ln{{\tilde L}} + 3.555 \cdot 10^{-4} \sqrt{{\tilde J}} + 1.520 \cdot 10^{-4} {\tilde J}^2 \ln{{\tilde L}} \\
& + 5.196 \cdot 10^{-4} {\tilde J}^2 - 9.378 \cdot 10^{-4} {\tilde J} \ln{{\tilde L}} - 1.547 \cdot 10^{-3} {\tilde J} \\
& + 0.002 \ln{{\tilde L}} + 1.643 \cdot 10^{-5} \ln{{\tilde J}} + 0.020
\end{align*}

\begin{align*}
a_3({\tilde L}, {\tilde J})_{LR} =\ & + 3.214 \cdot 10^{-4} \sqrt{{\tilde L}} \sqrt{{\tilde J}} - 8.363 \cdot 10^{-4} \sqrt{{\tilde L}} {\tilde J}^2 + 2.905 \cdot 10^{-3} \sqrt{{\tilde L}} {\tilde J} - 7.033 \cdot 10^{-3} \sqrt{{\tilde L}} \\
& + 2.348 \cdot 10^{-4} {\tilde L}^2 \sqrt{{\tilde J}} - 7.590 \cdot 10^{-4} {\tilde L}^2 {\tilde J}^2 + 1.648 \cdot 10^{-3} {\tilde L}^2 {\tilde J} - 1.763 \cdot 10^{-4} {\tilde L}^2 \\
& - 5.384 \cdot 10^{-4} {\tilde L} \sqrt{{\tilde J}} + 1.919 \cdot 10^{-3} {\tilde L} {\tilde J}^2 - 5.936 \cdot 10^{-3} {\tilde L} {\tilde J} + 5.277 \cdot 10^{-3} {\tilde L} \\
& - 3.766 \cdot 10^{-4} \sqrt{{\tilde J}} + 9.651 \cdot 10^{-4} {\tilde J}^2 - 7.959 \cdot 10^{-3} {\tilde J} + \frac{6.054 \cdot 10^{-6}}{{\tilde J}} \\ 
& + \frac{6.498 \cdot 10^{-5}}{{\tilde L}} + 5.780 \cdot 10^{-3}
\end{align*}


\begin{align*}
a_1({\tilde L}, {\tilde J})_{Poly-2} =\ & 0.023 {\tilde L}^2 - 6.087 \cdot 10^{-4} {\tilde L} {\tilde J} - 0.048 {\tilde L} - 3.062 \cdot 10^{-4} {\tilde J}^2 - 1.191 \cdot 10^{-4} {\tilde J} + 1.018
\end{align*}

\begin{align*}
a_2({\tilde L}, {\tilde J})_{Poly-2} =\ & 0.012 {\tilde L}^2 + 3.060 \cdot 10^{-4} {\tilde L} {\tilde J} - 0.030 {\tilde L} - 3.907 \cdot 10^{-4} {\tilde J}^2 + 0.006 {\tilde J} + 0.008
\end{align*}

\begin{align*}
a_3({\tilde L}, {\tilde J})_{Poly-2} =\ & 4.955 \cdot 10^{-3} {\tilde L}^2 - 1.255 \cdot 10^{-3} {\tilde L} {\tilde J} - 0.006 {\tilde L} + 1.388 \cdot 10^{-3} {\tilde J}^2 - 0.008 {\tilde J} + 0.005
\end{align*}

\begin{align*}
a_1({\tilde L}, {\tilde J})_{GP} =\ & 0.948 + \frac{4373.484}{{\tilde L}} - \frac{10666.073}{{\tilde L} + \frac{2}{{\tilde J}}} + \frac{0.693}{{\tilde J}} + \frac{78.987}{{\tilde L} + \frac{{\tilde J}}{{\tilde L}}} + \frac{6218.963}{{\tilde L} + \frac{\ln {\tilde L}}{{\tilde J}}}
\end{align*}

\begin{align*}
a_2({\tilde L}, {\tilde J})_{GP} =\ & 126757.788 - \frac{159623.632}{{\tilde L}} + \frac{21868.106}{{\tilde L} - \frac{6.086}{{\tilde J}}} + \frac{137755.287}{{\tilde L} + \frac{1}{{\tilde J}}} + \frac{797441.187}{-6.086 + \frac{1}{{\tilde L}{\tilde J}}} \\
& - \frac{840.950}{-0.197 + \frac{1}{{\tilde L}{\tilde J}}} + \frac{816.101}{{\tilde J}} - 7.738 \cdot 10^{-7} {\tilde L} {\tilde J} - \frac{26.366}{2.836 + {\tilde J}} - \frac{793.962}{- \frac{6.086}{{\tilde L}} + {\tilde J}}
\end{align*}

\begin{align*}
a_3({\tilde L}, {\tilde J})_{GP} =\ & -0.023 + \frac{49837.925}{{\tilde L}} + 1.312 \cdot 10^{-5} {\tilde L} + \frac{427.181}{\frac{3}{{\tilde L}} + {\tilde L}} - \frac{50263.596}{{\tilde L} + \frac{1}{{\tilde L}{\tilde J}}} \\
& - \frac{43.807}{{\tilde J}} - 3.182 \cdot 10^{-4} {\tilde J} + 3.248 \cdot 10^{-7} {\tilde L} {\tilde J} + \frac{44.804}{{\tilde J} + \frac{{\tilde J}}{{\tilde L}}}
\end{align*}

\twocolumngrid

\bibliographystyle{apsrev4-1}
\bibliography{bibliography_abbreviated_fixed}

\end{document}